\pdfoutput=1
\documentclass[a4paper,fleqn,usenatbib]{article}

\usepackage{newtxtext,newtxmath}
\usepackage[utf8]{inputenc}
\usepackage{amsmath}
\usepackage{abbreviations,natbib}

\usepackage{graphicx}
\usepackage[caption=false]{subfig}
\usepackage{hyperref}
\usepackage{color}
\usepackage{nameref}
\usepackage{wrapfig}
\usepackage{authblk}
\usepackage[a4paper,margin=2.5cm]{geometry}

\newcommand{\vecu}{\textbf{u}}
\newcommand{\vecupar}{\textbf{u}_{\|}}
\newcommand{\vecuperp}{\textbf{u}_\perp}
\newcommand{\upar}{u_{\|}}

\newcommand{\barvecuperpg}{\bar{\textbf{u}}_{\perp g}}
\newcommand{\baruperpg}{\bar{u}_{\perp g}}
\newcommand{\barvecu}{\bar{\textbf{u}}}

\newcommand{\barupar}{\bar{u}_\parallel}
\newcommand{\vecx}{\textbf{x}}
\newcommand{\vecR}{\textbf{R}}
\newcommand{\vecE}{\textbf{E}}
\newcommand{\vecD}{\textbf{D}}
\newcommand{\vecv}{\textbf{v}}
\newcommand{\vecB}{\textbf{B}}
\newcommand{\vecb}{\textbf{b}}
\newcommand{\geoG}{\mathcal{G}}
\newcommand{\geoL}{\mathcal{L}}
\newcommand{\geoV}{\mathcal{V}}

\newcommand{\vecuperpg}{\vecu_{\perp g}}
\newcommand{\uperpg}{u_{\perp g}}
\DeclareMathAlphabet\mathbfcal{OMS}{cmsy}{b}{n}
\begin{document}
\title{A coupled guiding center-Boris particle pusher for magnetized plasmas in compact-object magnetospheres}
\author[1]{F. Bacchini\thanks{E-mail: fabio.bacchini@kuleuven.be}}
\author[2,3]{B. Ripperda\thanks{Joint Princeton/Flatiron Postdoctoral Fellow}}
\author[2]{A.A. Philippov}
\author[3]{K. Parfrey}
\affil[1]{Centre for mathematical Plasma Astrophysics, Department of Mathematics, KU Leuven, Celestijnenlaan 200B, B-3001 Leuven, Belgium}
\affil[2]{Center for Computational Astrophysics, Flatiron Institute, 162 Fifth Avenue, New York, NY 10010, USA}
\affil[3]{Department of Astrophysical Sciences, Peyton Hall, Princeton University, Princeton, NJ 08544, USA}
\renewcommand\Authands{ and }

\label{firstpage}
\maketitle

\begin{abstract}
We present a novel numerical scheme for simulating the motion of relativistic charged particles in magnetospheres of compact objects, typically filled with highly magnetized collisionless plasmas. The new algorithm is based on a dynamic switch between the full system of equations of motion and a guiding center approximation. The switch between the two formulations is based on the magnetization of the plasma particles, such that the dynamics are accurately captured by the guiding center motion even when the gyro-frequency is under-resolved by the time step. For particles with a large gyro-radius, due to acceleration in, e.g., reconnecting current sheets, the algorithm adaptively switches to solve the full equations of motion instead. The new scheme is directly compatible with standard Particle-in-Cell codes, and is readily applicable in curved spacetimes via a dedicated covariant formulation. We test the performance of the coupled algorithm by evolving charged particles in electromagnetic configurations of reconnecting current sheets in magnetized plasma, obtained from special- and general-relativistic Particle-in-Cell simulations. The new coupled pusher is capable of producing highly accurate particle trajectories even when the time step is many orders of magnitude larger than the gyro-period, substantially reducing the restrictions of the temporal resolution.
\end{abstract}


\section{Introduction}
\label{sec:intro}
Magnetospheres of compact objects, such as black holes (BHs) and neutron stars (NSs), are typically filled with highly magnetized collisionless plasma. The collective plasma dynamics is best studied with a kinetic description governed by the Vlasov-Maxwell equations. The Particle-in-Cell (PiC) algorithm is the most widely employed method to solve the kinetic equations, owing its success to its simplicity, reliability, and remarkable performance on parallel architectures. Special-relativistic (SR) PiC simulations are used to study the dynamics of magnetized relativistic plasma, and have illuminated the physics of fundamental kinetic processes such as particle acceleration in shocks (\citealt{spitkovsky2008,sironispitkovsky2009a,sironispitkovsky2009b}), magnetic reconnection (\citealt{guo2014,sironispitkovsky2014,werneruzdensky2017,werner2018,hakobyan2019}), and turbulence (\citealt{comissosironi2018,comissosironi2019,zhdankin2018,zhdankin2019}). More recently, PiC simulations have been applied to model the plasma magnetospheres of compact objects, resulting in first-principles models of coherent radio emission from non-stationary reconnection (\citealt{philippov2019}) and pair discharges (\citealt{philippov2020}), $\gamma$-ray emission from reconnecting current sheets (\citealt{cerutti2016,kalapotharakos2018,philippovspitkovsky2018}), and X-ray emission in magnetar magnetospheres (\citealt{chenbeloborodov2014}). General-relativistic (GR) PiC simulations have investigated the dynamics of plasmas in curved spacetime, showing that dragging of inertial frames is an essential ingredient for pair creation at the pulsar's polar cap (\citealt{philippov2015b}), and unveiling the dynamics of pair discharges and jet launching in BH magnetospheres (\citealt{levinsoncerutti2018,chenyuan2019,parfrey2019,crinquand2020}). 

Collisionless plasma magnetospheres of compact objects typically display a very large separation between the global spatial and temporal scales (e.g.,\ the size of the object and its rotational frequency) and the typical dynamic scales of the particle motion (e.g.,\ the gyro-radius $\rho_C$ and the gyro-frequency $\Omega_C$). This imposes constraints on PiC methods, reducing the parameter range that can be realistically probed with simulations. Furthermore, in compact-object magnetospheres where gravity may play an important role, it is imperative to include the effects of curved spacetime, further complicating the solution algorithms. For highly magnetized magnetospheres the scale separation can become extreme, with gyro-radii $\sim10^{10}$ times smaller than the system size. This difference can be further enhanced by fast (on time scales of order $10^{-14}$ s near surfaces of NSs) synchrotron losses, pushing the particles to the zeroth Landau level. To mitigate the problem, a rescaling of the system's physical parameters (e.g.,\ to lower magnetic field strengths) is typically employed, artificially decreasing the scale separation. 

A compromise is provided by applying a guiding center approximation (GCA; see e.g.,\ \citealt{northrop1961,northrop1963}) reducing the particle motion to that of its guiding center. In the GCA equations, the particle dynamics is averaged over its gyro-motion, and therefore no constraints related to particle gyration are imposed on the time step. This approach has been successfully applied to study test particle motion in the solar corona (e.g.\ \citealt{rosdhalgalsgaard2009,gordovskyy2010,gordovskyybrowning2011a,gordovskyybrowning2011b,gordovskyy2014,zhou2015,pinto2016,zhou2016,ripperda2017a,ripperda2017b,threlfall2017}), in the Earth's magnetotail (e.g.\ \citealt{birn2004} and references therein; \citealt{leroy2019}), and has been used in PiC simulations of pair discharges (\citealt{philippov2020}). However, the GCA is based on the assumption that the particle gyro-radius is much smaller than the typical length scale of electromagnetic fluctuations. This implies that particles whose gyro-radius increases significantly, e.g.,\ due to acceleration or a local decrease of the magnetic field strength, may not be accurately described by the GCA equations. This scenario is expected in compact-object magnetospheres, where particles mainly travel along magnetic field lines with negligible gyro-radius, but may get accelerated and gain significant transverse momentum in reconnecting current sheets. The GCA alone is therefore inadequate to accurately model the acceleration of particles in such situations.

In this paper, we present a reliable particle pusher that both accurately captures particle dynamics when its gyro-frequency is significantly under-resolved by the time step, and when its gyro-radius becomes large due to acceleration in, e.g., reconnecting current sheets.We develop a novel algorithm that couples a leapfrog GCA solver with a standard Boris (\citealt{boris1970}) scheme, and switches between the two approaches according to criteria based on the magnetization of the plasma particles. The coupled method is also applicable in curved spacetimes, via a dedicated covariant formulation. The paper is organized as follows: in Section \ref{sec:theory} we review the relativistic equations of motion for charged particles in electromagnetic fields, and the GCA equations for the motion of the guiding center. We discuss the numerical methods employed for the solution of the equations of motion, presenting two leapfrog-like numerical schemes for the solution of the SR-GCA equations, and a method for the solution of the GR-GCA system. In Section \ref{sec:scheme} we present the new coupled Boris-GCA pusher, providing the necessary steps for the construction and implementation of the algorithm. We also present a GR version of the scheme which is applicable to the motion of charged particles in strong gravitational fields. In Section \ref{sec:tests}, we test the new coupled pusher by simulating the motion of charged particles in an isolated reconnecting current sheet in flat spacetime and a current sheet formed in a BH magnetosphere. Finally, in Section \ref{sec:conclusions} we summarize our results and present our main conclusions.

\section{Equations of motion and numerical methods for charged particles in electromagnetic fields}
\label{sec:theory}

In this Section we briefly review the theoretical framework describing the motion of charged particles in electromagnetic fields. We consider both the case of motion in flat spacetime (i.e. in the absence of gravity) and in curved spacetimes. In each case we first present the relevant equations of motion, and then describe the numerical algorithms we employ to solve them.

\subsection{Full system of equations in flat spacetime}
\label{sec:sreom}
The evolution equations for the particle position, $\vecx$, and 4-velocity, $\vecu$, read
\begin{equation}
 \frac{d\vecx}{dt} = \frac{\vecu}{\Gamma},
 \label{eq:dxdt}
\end{equation}
\begin{equation}
 \frac{d\vecu}{dt} = \frac{q}{m}\left(\vecE+\frac{\vecu}{c\Gamma}\times\vecB\right),
 \label{eq:dudt}
\end{equation}
where $\Gamma=\sqrt{1+u^2/c^2}$ is the Lorentz factor, $c$ is the speed of light, and $q$ and $m$ are the particle charge and mass. Note that we refer to $\vecu=(u_1,u_2,u_3)$ as the 4-velocity, for simplicity, although this is in fact only the spatial part of the 4-velocity vector. Equations \eqref{eq:dxdt}--\eqref{eq:dudt} are typically approximated with a second-order time discretization, where $\vecx$ and $\vecu$ are staggered in time. The solution of equation \eqref{eq:dudt} is usually carried out with a structure-preserving algorithm such as the Boris approach (\citealt{boris1970}), which possesses desirable qualities (e.g.,\ the conservation of phase-space volume). Several other strategies have been presented over the years, to address specific numerical requirements (see \citealt{ripperda2018a} and references therein for a comprehensive comparison of particle pushers). All approaches are however constrained on the time step, which must be typically kept to a value comparable to the gyro-period $t_C=2\pi m c\Gamma/(|q|B)$. This condition may become extremely demanding for strong magnetic fields, as is often the case for compact-object magnetospheres. Hereafter, we will refer to equations \eqref{eq:dxdt}--\eqref{eq:dudt} as the "special-relativistic equations of motion" (\hyperref[eq:dxdt]{SREOM}), for clarity.

\subsection{Guiding Center Approximation in flat spacetime}
\label{sec:srgca}
In the guiding center formalism, the guiding center position $\vecR$ is evolved in place of $\vecx$ (see e.g.,\ \citealt{northrop1961,northrop1963}). The particle 4-velocity $\vecu$ is split as $\vecu=\vecupar+\vecu_{\perp d} + \vecuperpg$, where
\begin{equation}
 \vecupar = (\vecu\cdot\vecb^*)\vecb^*,
\end{equation}
is defined with respect to $\vecb^*$, a unit vector parallel to the magnetic field in a frame of reference where $\vecE$ and $\vecB$ are parallel (\citealt{beklemishevtessarotto1999}). This frame moves with the velocity
\begin{equation}
\vecv_E = \frac{\textbf{w}_E}{2w_E^2/c^2}\left(1-\sqrt{1-4w_E^2/c^2}\right),
\end{equation}
where $\textbf{w}_E=c\vecE\times\vecB/(E^2+B^2)$, defining an associated Lorentz factor $\kappa=1/\sqrt{1-v_E^2/c^2}$. The drift component $\vecu_{\perp d}$ and the gyration component $\vecuperpg$ describe the guiding center drift across magnetic field lines and the particle gyro-motion (at the cyclotron frequency $\Omega_C=|q|B/(mc\Gamma)$) in the plane perpendicular to $\vecb^*$, respectively.

Note that both $\vecb^*$ and $\vecb=\vecB/B$, the unit vector in the direction of the lab-frame magnetic field, lie in the plane perpendicular to $\vecE\times\vecB$ and hence to $\vecv_E$ (\citealt{beklemishevtessarotto1999}). The two unit vectors differ by a rotation of $\vecb^*=\xi\vecE_\perp + \eta\vecB$, where $\vecE_\perp=\vecE-(\vecE\cdot\vecB)\vecB/B^2 = {\vecB\times(\vecE\times\vecB)/B^2}$, $\xi=\eta\Psi$, $\eta=\left(B\sqrt{\Psi^2E^2_\perp/B^2+1}\right)^{-1}$, and $\Psi=E'(B^2-B'^2)/(B'E_\perp^2)$. Here, $E'$ and $B'$ are the magnitudes of the electric and magnetic field in the reference frame moving with $\vecv_E$,
\begin{equation}
 E' = \frac{\vecE\cdot\vecB}{B'},
\end{equation}
\begin{equation}
 B' = \sqrt{\frac{1}{2}\left(B^2-E^2+(B^2+E^2)\sqrt{1-4 w_E^2/c^2}\right)}.
\end{equation}
In the limit $\vecE\cdot\vecB\rightarrow0$ (alternatively, $\vecE\cdot\vecB/B \ll E_{\perp}$), it can be shown that $\vecb^*\rightarrow(\vecB+\vecE_\perp(\vecE\cdot\vecB)/B'^2)/B$, i.e.,\ $\vecb^*$ and $\vecb$ differ by a first-order correction proportional to the parallel electric field $(\vecE\cdot\vecB)/B$. This correction is equivalent to the purely relativistic drift term in  \cite{northrop1963,ripperda2018a}, i.e.\ $\upar\vecb^*\rightarrow\upar\vecb + \vecu_\mathrm{rel}$, where
\begin{equation}
 \vecu_\mathrm{rel} = \upar(\vecE\cdot\vecB)\frac{\vecb}{B'^2}\times\frac{\vecE\times\vecB}{B^2}
\end{equation}
and $B'=B\sqrt{1-E_\perp^2/B^2}$ in this limit. The equations above simplify significantly when $\vecE\cdot\vecB=0$ exactly, since $\xi=0$ and $\eta=1/B$, yielding $\vecb^*=\vecb$. In such a case, $\vecupar$ lies exactly along the direction of $\vecB$ and the relativistic drift term vanishes. Additionally, for $\vecE\cdot\vecB=0$ the velocity $\vecv_E$ reduces to a familiar drift velocity with magnitude $v_E=c E/B$ (if $E<B$) or $v_E=c B/E$ (if $E>B$), which corresponds to vanishing electric (if $E<B$) or magnetic (if $E>B$) field in the frame moving with $\vecv_E$. In all our experiments in the following Sections, the regimes where the GCA is employed correspond to the plasma-filled regions, where $\vecE\cdot\vecB/B \ll E_{\perp}$ is fulfilled. Therefore, in what follows we will assume that the velocity can be decomposed as $\vecu=\upar\vecb+\vecv_E\Gamma+\vecuperpg$, with $\vecv_E=c\vecE\times\vecB/B^2$, for simplicity. We verified that this assumption has no practical effect on the results presented here. The complete expressions above for $\vecb^*$ and $\vecv_E$ only differ from our approximated form by terms given as functions of $\vecE$ and $\vecB$; hence they can be evolved in numerical schemes such as those presented in the following Sections without significant extra effort.

While the parallel component $\upar$ is directly evolved in time, the  gyration component $\uperpg$ is instead determined by the evolution of the magnetic moment $\mu=m \uperpg^2/(2B\kappa)$, which is an adiabatic invariant of the motion (to first order) in the GCA equations. The validity of the GCA relies (among others) on the assumption that the particle gyro-radius $\rho_C=mc\uperpg/(|q|B)$ is much smaller than the length scales characterizing spatial variations of the electromagnetic fields. Therefore, numerical methods based on the GCA require careful checks on the validity of this assumption, which may be violated when particles experience strong acceleration (e.g.,\ in reconnection regions, around X-points, where $E>B$) in the direction perpendicular to $\vecB$, resulting in non-negligible gyro-motion.

The GCA equations are derived by means of an expansion of the \hyperref[eq:dxdt]{SREOM} system in powers of $\rho_C/L$ (where $L$ is the characteristic length scale of variation of electromagnetic fields), resulting in drift terms (perpendicular to magnetic field lines) proportional to spatial and temporal variations of $E$ and $B$ (\citealt{vandervoort1960,northrop1961,northrop1963}). In the expansion, $\vecuperp$ is split into several components of which the $\vecv_E$ drift represents the only zeroth-order term; if $E\ll B$ and temporal derivatives of the electromagnetic fields are neglected (under the assumption that particle dynamics takes place on much faster time scales than the electromagnetic field evolution), two additional first-order terms arise, namely the curvature and $\nabla B$ drifts. The curvature drift is given by
\begin{equation}
 \vecv_c = \frac{mc\kappa^2}{qB}\vecb\times\left[ \frac{\upar^2}{\Gamma}(\vecb\cdot\nabla)\vecb + \upar (\vecv_E\cdot\nabla)\vecb\right],
 \label{eq:curvdrift}
\end{equation}
and it can be shown that the relative magnitude of $\vecv_c$ and $\vecv_E$ is
\begin{equation}
 \frac{v_c}{v_E} \sim \frac{\rho_{C}}{L},
\end{equation}
where $\rho_{C}=mc\Gamma/(|q|B)$ is the typical gyro-radius of particles traveling with Lorentz factor $\Gamma$. It can be seen that, as the particle magnetization (parametrized by $L/\rho_{C}$) increases, the curvature drift becomes progressively less important than the $\vecE\times\vecB$ drift, which instead does not scale with $\rho_{C}$. Similarly, in the evolution equation for $\upar$ the expansion introduces zeroth-order terms (given by parallel electric fields) and first-order curvature acceleration terms. Again, the relative magnitude of these terms scales such that, for large magnetizations, the effects of curvature are negligible.

The $\nabla B$ ("mirror") drift velocity $\vecv_{\nabla B}$ is proportional to $\mu$, and it can be shown that the relative magnitude of the mirror and curvature drift velocities scales as $v_{\nabla B}/v_c\sim v_{\perp g}^2/v_{\|}^2$, where $v_{\perp g}=\uperpg/\Gamma$ and $v_{\|}=\upar/\Gamma$. First-order mirror force terms, which enter in the evolution equation for $\upar$, scale similarly with respect to curvature terms as $v_{\perp g}^2/(v_E v_{\|})$. In practice, for moderately relativistic particles with $v_{\|}\sim c$, mirror effects can be neglected as long as $v_{\perp g}\ll c$ (which implies small $\mu$). We verify that in all our experiments this condition is respected at all times; this scenario is representative of physical situations where gyration has negligible importance in the overall particle dynamics (e.g.,\ in the case of compact-object magnetospheres). For these reasons, we choose to discard mirror drift and acceleration terms in the following equations. Note that, in principle, the condition for which mirror effects are negligible should be rigorously checked in numerical methods based on the GCA equations, to avoid large inaccuracies in the results.

By these arguments, in this work we choose to retain $\vecv_E$ and $\vecv_c$ as the only relevant drift terms (as well as curvature acceleration terms), bearing in mind that the latter may not play a significant role for scenarios in which the plasma magnetization is sufficiently high. Hence, we solve the equations of motion for this "reduced" GCA,
\begin{equation}
 \frac{d\vecR}{dt} = \frac{\vecupar}{\Gamma} + \vecv_E + \vecv_c,
 \label{eq:dRdt}
\end{equation}
\begin{equation}
 \frac{d\upar}{dt} = \frac{q}{m}E_{\|} + \upar\vecv_E\cdot(\vecb\cdot\nabla)\vecb + \Gamma\vecv_E\cdot(\vecv_E\cdot\nabla)\vecb,
 \label{eq:dupardt}   
\end{equation}
\begin{equation}
 \frac{d\mu}{dt} = 0,
 \label{eq:dmudt}   
\end{equation}
where $E_{\|} = \vecE\cdot\vecb$. The Lorentz factor here is calculated by averaging energy fluctuations over a gyro-period,
\begin{equation}
\Gamma=\kappa\sqrt{1+(\upar^2+\uperpg^2)/c^2}.
\label{eq:lfacgca}
\end{equation}
We will hereafter refer to equations \eqref{eq:dRdt}--\eqref{eq:dmudt} as the "special-relativistic GCA" (\hyperref[eq:dRdt]{SRGCA}) system for clarity. Note that neglecting mirror force terms in equation \eqref{eq:dupardt} may in principle introduce errors in the particle energy: in the case where $E=0$ and curvature terms are neglected, the particle energy as expressed by equation \eqref{eq:lfacgca} should remain constant, but since there is no mechanism of conversion of perpendicular velocity into parallel velocity, solving equation \eqref{eq:dmudt} for $\uperpg$ will introduce a deviation from exact energy conservation. This effect is in practice negligible as long as mirror forces (determined by $\mu$) are negligible, as discussed above. An alternative approach consists of avoiding the solution of equation \eqref{eq:dmudt} and keeping $\uperpg$ as a constant. In this case, energy is conserved exactly, at the expense of numerically altering $\mu$. In all our experiments this deviation from exact energy conservation does not significantly influence the results, since mirror forces remain negligible at all times. We therefore rely on the former approach, including the solution of equation \eqref{eq:dmudt}.

For compatibility with standard PiC codes, we wish to solve the system above with a leapfrog-like, second-order scheme. Here we present such a scheme consisting of a semi-implicit discretization of the equations of motion, where the position $\vecR$ is centered on integer time steps, while $\upar$ is centered on half steps. We analyze two subcases, namely the case where curvature effects are negligible and the case that includes curvature. In compact-object magnetospheres, $\mu$ is typically quickly brought to zero via synchrotron radiation. Therefore, in the algorithms presented below $\mu=0$ can be assumed for specific applications.

\subsubsection{Zero curvature case}
\label{sec:gcamethod1}
In cases where it is possible to neglect all gradient terms in the \hyperref[eq:dRdt]{SRGCA} system (e.g.,\ for highly magnetized plasmas where $\vecv_c$ and curvature acceleration terms become negligible), the equations of motion take the simple form
\begin{equation}
 \frac{d\vecR}{dt} = \frac{\vecupar}{\Gamma} + \vecv_E,
 \label{eq:dRdtcase1}
\end{equation}
\begin{equation}
 \frac{d\upar}{dt} = \frac{q}{m}E_{\|}.
 \label{eq:dupardtcase1}   
\end{equation}
These equations can be solved with a leapfrog-like approach which relies on the second-order discrete equations
\begin{equation}
 \frac{\vecR^{n+1}-\vecR^{n}}{\Delta t} = \frac{\upar^{n+1/2}}{2}\left(\frac{\vecb^n}{\Gamma(\vecR^n,\upar^{n+1/2})} + \frac{\vecb^{n+1}}{\Gamma(\vecR^{n+1},\upar^{n+1/2})} \right) + \frac{\vecv_E^n + \vecv_E^{n+1}}{2},
 \label{eq:dRdtcase1disc}
\end{equation}
\begin{equation}
 \frac{\upar^{n+1/2}-\upar^{n-1/2}}{\Delta t} = \frac{q}{m}E_{\|}^n,
 \label{eq:dupardtcase1disc}
\end{equation}
where we use a short-hand notation for quantities that depend on $\vecR$ only, e.g., $\vecb^n=\vecb(\vecR^n)$, $\vecv_E^n=\vecv_E(\vecR^n)$, etc. The Lorentz factor is a function of both $\vecR$ and $\upar$,
\begin{equation}
 \Gamma(\vecR,\upar) = \kappa(\vecR) \sqrt{1+(\upar^2+2\mu B(\vecR)\kappa(\vecR)/m)/c^2},
 \label{eq:lfacgca2}
\end{equation}
and $\mu$ is a conserved quantity across a time step. Equations \eqref{eq:dRdtcase1disc} and \eqref{eq:dupardtcase1disc} can be solved, at each time step, in a semi-implicit fashion. While equation \eqref{eq:dupardtcase1disc} can be solved explicitly, equation \eqref{eq:dRdtcase1disc} is nonlinearly implicit in the new position $\vecR^{n+1}$. At each time step, starting from $\vecR^n$ and $\vecupar^{n-1/2}$, we therefore apply the following solution procedure:
\begin{enumerate}
 \item Solve the 4-velocity equation \eqref{eq:dupardtcase1disc} explicitly,
 \begin{equation}
  \upar^{n+1/2} = \upar^{n-1/2} + \frac{q\Delta t}{m} E_{\|}^n.
 \end{equation}
 \item Solve equation \eqref{eq:dRdtcase1disc} for $\vecR^{n+1}$ iteratively, e.g.,\ with a fixed-point method. An appropriate initial guess is given by the previous value $\vecR^n$. At the generic $k$-th iteration, the nonlinear residual to be evaluated reads
 \begin{equation}
  H(\vecR^k) = \vecR^k-\vecR^n - \frac{\Delta t}{2}\upar^{n+1/2}\left(\frac{\vecb^n}{\Gamma(\vecR^n,\upar^{n+1/2})} + \frac{\vecb^k}{\Gamma(\vecR^k,\upar^{n+1/2})} \right) - \frac{\Delta t}{2}(\vecv_E^n + \vecv_E^k).
 \end{equation}
 In our experiments, 3-4 iterations are sufficient to reach convergence within an absolute iteration error of $10^{-7}$.
\end{enumerate}
This approach has been successfully applied in multi-dimensional simulations of pair discharges in pulsar polar caps (\citealt{philippov2020}), in a scenario where $\vecv_E$ acts as the predominant drift mechanism in the motion of particles.

\subsubsection{Nonzero curvature case}
\label{sec:gcamethod2}
If the magnetic field curvature cannot be neglected, it is necessary to include the curvature terms in the \hyperref[eq:dRdt]{SRGCA} system. Applying a second-order discretization yields
\begin{equation}
 \frac{\vecR^{n+1}-\vecR^{n}}{\Delta t} = \frac{\upar^{n+1/2}}{2}\left(\frac{\vecb^n}{\Gamma(\vecR^n,\upar^{n+1/2})} + \frac{\vecb^{n+1}}{\Gamma(\vecR^{n+1},\upar^{n+1/2})} \right) + \frac{\vecv_E^n + \vecv_E^{n+1}}{2} + \frac{\vecv_c(\vecR^n,\upar^{n+1/2}) + \vecv_c(\vecR^{n+1},\upar^{n+1/2})}{2},
 \label{eq:dRdtcase2disc}
\end{equation}
\begin{equation}
 \frac{\upar^{n+1/2}-\upar^{n-1/2}}{\Delta t} = \frac{q}{m}E_{\|}^n + \frac{\upar^{n-1/2}+\upar^{n+1/2}}{2} \vecv_E^n\cdot(\vecb^n\cdot\nabla)\vecb^n + \frac{1}{2}\left(\Gamma(\vecR^n,\upar^{n-1/2})+\Gamma(\vecR^n,\upar^{n+1/2})\right)\vecv_E^n\cdot(\vecv_E^n\cdot\nabla)\vecb^n,
 \label{eq:dupardtcase2disc}
\end{equation}
where the curvature drift is a function of both $\vecR$ and $\upar$,
\begin{equation}
 \vecv_c(\vecR,\upar) = \frac{mc\kappa^2(\vecR)}{qB(\vecR)}\vecb(\vecR)\times\left[ \frac{\upar^2}{\Gamma(\vecR,\upar)}(\vecb(\vecR)\cdot\nabla)\vecb(\vecR) + \upar (\vecv_E(\vecR)\cdot\nabla)\vecb(\vecR)\right].
\end{equation}
While the system \eqref{eq:dRdtcase2disc}--\eqref{eq:dupardtcase2disc} is now manifestly more complex than in the zero curvature case, the solution procedure remains analogous and can be carried out according to the following steps:
\begin{enumerate}
 \item Solve the 4-velocity equation \eqref{eq:dupardtcase2disc} explicitly, via
 \begin{equation}
  \upar^{n+1/2} = \left(\upar' + \frac{\Delta t}{2}\Gamma(\vecR^n,\upar^{n+1/2})\vecv_E^n\cdot(\vecv_E^n\cdot\nabla)\vecb^n \right)\left( 1-\frac{\Delta t}{2}\vecv_E^n\cdot(\vecb^n\cdot\nabla)\vecb^n\right)^{-1},
  \label{eq:dupardtcase2discsol}
 \end{equation}
 with
 \begin{equation}
  \upar' = \upar^{n-1/2}\left(1+\frac{\Delta t}{2}\vecv_E^n\cdot(\vecb^n\cdot\nabla)\vecb^n\right) + \frac{q\Delta t}{m}E_{\|}^n + \frac{\Delta t}{2}\Gamma(\vecR^n,\upar^{n-1/2})\vecv_E^n\cdot(\vecv_E^n\cdot\nabla)\vecb^n.
   \label{eq:duparprimedtcase2discsol}
 \end{equation}
 The Lorentz factor $\Gamma(\vecR^n,\upar^{n+1/2})$ needed in equation \eqref{eq:dupardtcase2discsol} can be calculated explicitly by substituting equation \eqref{eq:duparprimedtcase2discsol} into equation \eqref{eq:dupardtcase2discsol}. Then, by inverting equation \eqref{eq:lfacgca2}, we obtain a single, quadratic equation in $\Gamma(\vecR^n,\upar^{n+1/2})$. The Lorentz factor is then given by the only positive (larger than or equal to 1) solution of a second-order polynomial $k_1\Gamma^2 + k_2\Gamma + k_3 = 0$ with coefficients
 \begin{equation}
  k_1 = \frac{(\Delta t)^2}{4}\left(\vecv_E^n\cdot(\vecv_E^n\cdot\nabla)\vecb^n\right)^2 - \frac{c^2}{(\kappa^n)^2}\left(1 - \frac{\Delta t}{2}\vecv_E^n\cdot(\vecb^n\cdot\nabla)\vecb^n\right)^2,
 \end{equation}
 \begin{equation}
  k_2 = \upar'\Delta t \vecv_E^n\cdot(\vecv_E^n\cdot\nabla)\vecb^n,
 \end{equation}
 \begin{equation}
  k_3 = (\upar')^2 + (c^2+2\mu B^n\kappa^n/m)\left(1-\frac{\Delta t}{2}\vecv_E^n\cdot(\vecb^n\cdot\nabla)\vecb^n\right)^2.
 \end{equation}
 \item Solve equation \eqref{eq:dRdtcase2disc} for $\vecR^{n+1}$ iteratively. At the generic $k$-th iteration, the nonlinear residual to be evaluated reads
 \begin{equation}
  H(\vecR^k) = \vecR^k-\vecR^n - \frac{\Delta t}{2}\upar^{n+1/2}\left(\frac{\vecb^n}{\Gamma(\vecR^n,\upar^{n+1/2})} + \frac{\vecb^k}{\Gamma(\vecR^k,\upar^{n+1/2})} \right) - \frac{\Delta t}{2}(\vecv_E^n + \vecv_E^k) - \frac{\Delta t}{2}\left(\vecv_c(\vecR^n,\upar^{n+1/2}) + \vecv_c(\vecR^k,\upar^{n+1/2})\right).
 \end{equation}
 Again, 3-4 iterations usually suffice to reach convergence within an absolute iteration error of $10^{-7}$.
\end{enumerate}

\subsection{Full system of equations in curved spacetimes}
\label{sec:greom}
For compact-object environments where strong gravitational fields are present, spacetime curvature effects may become important, influencing the motion of particles. A general-relativistic formulation of the equations of motion for a charged particle in the presence of gravitational and electromagnetic fields reads
\begin{equation}
 \frac{dx^i}{dt} = \geoV^i,
 \label{eq:GRdxdt}
\end{equation}
\begin{equation}
 \frac{du_i}{dt} = \geoG_{i}+\geoL_i,
 \label{eq:GRdudt}
\end{equation}
where the 3+1 split of the Arnowitt-Deser-Misner (ADM) formalism (\citealt{arnowitt1959}) has been employed to define the evolution of the particle position $x^i$ and 4-velocity $u_i$ with respect to coordinate time $t$ (Roman indices run from 1 to 3). The right-hand side terms correspond to the particle motion and geodesic terms,
\begin{equation}
 \geoV^i = \alpha\frac{\gamma^{ij}u_j}{\Gamma} - \beta^i
\end{equation}
\begin{equation}
 \geoG_{i}=-\Gamma\partial_i\alpha+u_j\partial_i\beta^j-\alpha\frac{u_j u_k}{2\Gamma}\partial_i\gamma^{jk},
\end{equation}
and to the covariant Lorentz force,
\begin{equation}
 \geoL_i = \alpha\frac{q}{m}\left(\gamma_{ij}D^j+e_{ijk}\frac{\gamma^{jl}u_l}{\Gamma}B^k\right).
\end{equation}
Here, $\alpha$, $\beta^i$, and $\gamma_{ij}$ represent the metric functions (lapse, shift, and spatial 3-metric) of a Riemannian manifold $g_{\mu\nu}$ (Greek indices run from 0 to 3), with associated Levi-Civita pseudotensor $e_{ijk}=\sqrt{\gamma}\epsilon_{ijk}$. In this section we employ geometrized units such that $c=G=1$. The Lorentz factor in this formalism is defined as $\Gamma=\sqrt{1+\gamma^{ij}u_i u_j}$, and the electric and magnetic fields are constructed by projecting the electromagnetic field tensor $F^{\mu\nu}$,
\begin{equation}
 D^i=\alpha F^{0i},
\end{equation}
\begin{equation}
 B^i=\frac{1}{2}e^{ijk} F_{jk}.
\end{equation}
For clarity, we will refer to equations \eqref{eq:GRdxdt}--\eqref{eq:GRdudt} as the "general-relativistic equations of motion" (\hyperref[eq:GRdxdt]{GREOM}) from now on.

The highly nonlinear character of the \hyperref[eq:GRdxdt]{GREOM} system does not allow for the application of explicit, leapfrog-like discretization schemes. However, the motion along geodesics combined with the Lorentz force still demands for the accurate preservation of phase-space trajectories. For this reason, an appropriate choice for the solution of the system above is the implicit midpoint rule (IMR), which is a symplectic method (see e.g.,\ \citealt{hairer}). Alternative numerical schemes have been proposed, e.g.,\ based on a Hamiltonian formulation that preserves energy exactly during the simulation (\citealt{bacchini2018a,bacchini2019a}). However, the Lorentz force and the geodesic motion may impose dynamics on vastly different time scales (e.g.,\ for particles whose gyro-motion is much faster than the motion along geodesics), resulting in the \hyperref[eq:GRdxdt]{GREOM} system becoming stiff, which complicates the solution procedure.

An alternative approach consists of Strang splitting the equations above, by discretizing the geodesic motion with a symplectic scheme (e.g.,\ IMR), while applying the electromagnetic push of the Lorentz force in a local inertial reference frame (in which spacetime curvature can be ignored for instantaneous pointwise interactions). The projection to and from the local frame is obtained via a set of orthonormal tetrads. In this frame, the Lorentz force in the 4-velocity update simply takes the standard special-relativistic form (equation \eqref{eq:dudt}), and can be solved explicitly, e.g.,\ with a Boris algorithm (see \citealt{parfrey2019,crinquand2020}; the method will be presented in detail in a future publication (Parfrey et al. 2020, in prep.)). The overall scheme retains second-order accuracy. The advance of the particle position $\vecx=(x^1,x^2,x^3)$ and 4-velocity $\vecu=(u_1,u_2,u_3)$ from time level $n$ to $n+1$ (note that the variables are non-staggered) can be summarized in the following substeps:
\begin{enumerate}
 \item First half-push: the 4-velocity $\vecu^n$ is first projected to a local inertial frame, evaluating the metric functions at $\vecx^n$. Then, an updated 4-velocity $\vecu'$ is obtained by applying a Boris push on $\vecu^n$ over half a time step. The electromagnetic fields needed for the update are calculated by projecting to the same local frame. After the update, $\vecu'$ is projected back to the global coordinate basis.
 
 \item Position update and geodesic motion: the particle position and 4-velocity are evolved in the global coordinates via the 3+1 split geodesic equations, updating $\vecu'$ with the contribution from geometric sources. The equations of motion are discretized with an IMR scheme,
 \begin{equation}
  \begin{aligned}
   \frac{\vecx^{n+1}-\vecx^n}{\Delta t} & = \mathbfcal{V}(\bar{\vecx},\bar{\vecu}) \\
   \frac{\vecu''-\vecu'}{\Delta t} & = \mathbfcal{G}(\bar{\vecx},\bar{\vecu})
  \end{aligned},
 \end{equation}
 where $\bar{\vecx}=(\vecx^{n+1}+\vecx^n)/2$ and $\bar{\vecu}=(\vecu''+\vecu')/2$. The system above is implicit in $\vecx^{n+1}$ and $\vecu''$, and therefore requires a nonlinear iteration, which usually converges quickly provided that the metric functions do not introduce any stiffness. In our experiments, 3-4 fixed-point iterations with $\vecx^n$ and $\vecu'$ as initial guesses are generally sufficient to reach convergence within an absolute iteration error of $10^{-7}$.

 \item Second half-push: the 4-velocity $\vecu''$ is projected to a local inertial frame, and a second Boris push is carried out over half a time step, bringing $\vecu''$ to $\vecu^{n+1}$ (with metric and fields calculated at $\vecx^{n+1}$).
\end{enumerate}

We note that the half pushes (steps 1. and 3. above) are applied in a local inertial frame only for the sake of simplicity; in fact, a covariant Boris-like push in the global coordinate basis (without the need to project to a local frame) can still be applied explicitly, although it involves more evaluations of the metric functions. This option could be considered for experiments in spacetimes where calculating a suitable tetrad basis is non-straightforward, or too expensive.

\subsection{Guiding Center Approximation in curved spacetime}
\label{sec:grgca}
When spacetime curvature is considered, a formal treatment of the gyrokinetic approximation leading to the GCA equations becomes much more complicated than the flat-spacetime case presented in Section \ref{sec:srgca} (see e.g.,\ \citealt{beklemishevtessarotto1999,beklemishevtessarotto2004,cremaschini2008}). Here, we present a substantially simplified version of the covariant GCA, constructed such that the resulting equations can be solved with an approach similar to that discussed for the \hyperref[eq:GRdxdt]{GREOM} system in Section \ref{sec:greom}. In this way we ensure that the two systems of equations can be employed together in a coupled numerical scheme. We consider the following equations of motion,
\begin{equation}
 \frac{dR^i}{dt} = \tilde{\geoV}^i,
 \label{eq:GRdRdt}
\end{equation}
\begin{equation}
 \frac{d\upar}{dt} = (\geoG_{i}+\geoL_i)b^i,
 \label{eq:GRdupardt}
\end{equation}
\begin{equation}
 \frac{d\mu}{dt} = 0,
 \label{eq:GRdmudt}
\end{equation}
where $R^i$ is the guiding center position, and
\begin{equation}
 \tilde{\geoV}^i = \alpha\left(\frac{\upar\gamma^{ij} b_j}{\Gamma} + \gamma^{ij}v_{E,j}\right) - \beta^i.
 \label{eq:geoV}
\end{equation}
In these equations, $u_i$ has been split into the usual parallel, drift, and gyration components under the same assumptions outlined in Section \ref{sec:srgca} (i.e.\ nearly perpendicular electric and magnetic fields), $u_i = \upar b_i + v_{E,i}\Gamma + u_{\perp g,i}$, where $b_i=B_i/B$ (with $B=\sqrt{B_i B^i}$) is the unit vector in the direction parallel to the magnetic field. Here, we again assume the $\vecD\times\vecB$ drift as the predominant cross-field motion, and neglect all other drift terms. We therefore take the drift velocity
\begin{equation}
 v_{E,i} = \frac{e_{ijk}D^j B^k}{B_l B^l},
\end{equation}
defining a drifting reference frame with associated Lorentz factor $\kappa=1/\sqrt{1-\gamma^{ij}v_{E,i}v_{E,j}}$. With these definitions, equation \eqref{eq:GRdRdt} describes the motion of the guiding center along the parallel and drift directions, and equation \eqref{eq:GRdupardt} describes the change in magnitude of the parallel 4-velocity component. The time-variation of the parallel 4-velocity is given by the projection of both the geodesic (apparent) force term $\geoG_{i}$ and the Lorentz force $\geoL_i$. In flat spacetime, the geodesic term vanishes and the SR limit of the \hyperref[eq:dRdt]{SRGCA} system is retrieved (see Section \ref{sec:srgca}). Finally, equation \eqref{eq:GRdmudt} expresses the conservation of the magnetic moment $\mu=m\gamma^{ij}u_{\perp g,i}u_{\perp g,j}/(2B\kappa)$. This directly provides $\uperpg \equiv \sqrt{\gamma^{ij}u_{\perp g,i}u_{\perp g,j}}$, analogously to the special-relativistic case. The Lorentz factor can be calculated from the GCA variables as
\begin{equation}
 \Gamma = \kappa\sqrt{1+ \upar^2 + 2\mu B\kappa/m}.
 \label{eq:grgcalfac}
\end{equation}
We will refer to equations \eqref{eq:GRdRdt}--\eqref{eq:GRdmudt} as the "general-relativistic GCA" (\hyperref[eq:GRdRdt]{GRGCA}) system from here on. The choice of only including the $\vecD\times\vecB$ drift may not be applicable for scenarios where, for example, accurate calculation of particle energy losses to curvature radiation is important. However, if radiation losses are negligible, curvature drifts as well as other drift mechanisms will be of second-order importance compared to $\vecD\times\vecB$ drifts. Our tests show that our simplified approach already provides satisfactory results for particle motion in the upstream of current sheets in a BH magnetosphere (see Section \ref{sec:tests}).

The numerical solution of the \hyperref[eq:GRdRdt]{GRGCA} system can be carried out with a Strang split approach similar to the scheme presented in the previous Section. Once again, we apply the contribution of the Lorentz force to the 4-velocity update in a local inertial reference frame, via a set of orthonormal tetrads. The position update and the geodesic push are computed iteratively with an IMR scheme. The resulting algorithm for the update of $R^i$ and $\upar$, applied between time levels $n$ and $n+1$, is composed of steps below. For simplicity, in the following we will employ the usual notation for 3-vectors $\vecR=(R^1,R^2,R^3)$, $\vecu=(u_1,u_2,u_3)$, etc.
\begin{enumerate}
 \item First half-push: the parallel 4-velocity $\vecupar^n$ is projected to a local inertial frame, evaluating the metric functions at $\vecR^n$. In this frame, we update $\upar^n$ to $\upar'$ over half a time step, evaluating the acceleration determined by the parallel electric field $D^n_\parallel = \vecD^n\cdot\vecb^n$. Here, we again rely on a short-hand notation for quantities that depend on $\vecR$ only, e.g.,\ $\vecD^n = \vecD(\vecR^n)$, etc. The updated parallel 4-velocity $\vecupar'=\upar'\vecb^n$ is then projected back to the global coordinate basis.
    
 \item Position update and geodesic motion: the particle position and 4-velocity are evolved in the global coordinates via the 3+1 split geodesic equations. The equations of motion are discretized with an IMR scheme,
 \begin{equation}
  \begin{aligned}
   \frac{\vecR^{n+1}-\vecR^n}{\Delta t} & = \tilde{\mathbfcal{V}}(\bar{\vecR},\barupar) \\
   \frac{\upar''-\upar'}{\Delta t} & = \mathbfcal{G}(\bar{\vecR},\barvecu)\cdot\vecb(\bar{\vecR}) \\
  \end{aligned},
 \end{equation}
 where the mid-point iteration variables are $\bar{\vecR}=(\vecR^{n+1}+\vecR^n)/2$ and $\barupar=(\upar''+\upar')/2$. The system above is implicit in $\vecR^{n+1}$ and $\upar''$, and can be solved iteratively. At each iteration step, in order to evaluate the $\mathbfcal{G}$ term, we reconstruct the full velocity vector as $\barvecu = \barupar\vecb(\bar{\vecR}) + \vecv_E(\bar{\vecR})\Gamma(\bar{\vecR},\barupar) + \barvecuperpg $. The gyration part of the 4-velocity is updated in magnitude, $\baruperpg=\sqrt{2\mu\kappa(\bar{\vecR})B(\bar{\vecR})/m}$, but kept fixed in direction, i.e.\ $\barvecuperpg=\baruperpg\vecb_{\perp g}^n$ where $\vecb_{\perp g}^n = \vecuperpg^n/\uperpg^n$. The $\tilde{\mathbfcal{V}}$ term and the Lorentz factor are given by equations \eqref{eq:geoV} and \eqref{eq:grgcalfac} respectively, as functions of $\vecR$ and $\upar$. Similarly to the iterative scheme employed for the \hyperref[eq:GRdxdt]{GREOM} system, 3-4 fixed-point iterations with $\vecR^n$ and $\upar'$ as initial guesses suffice to reach convergence within an absolute iteration error of $10^{-7}$, in all our experiments.
 
 \item Second half-push: the updated parallel 4-velocity $\vecupar''$ is again projected to a local inertial frame, evaluating the metric functions at $\vecR^{n+1}$. The updated parallel 4-velocity is calculated in this frame by adding the contribution of the parallel electric field $D^{n+1}_\parallel=\vecD^{n+1}\cdot\vecb^{n+1}$. Finally,  $\vecupar^{n+1}=\upar^{n+1}\vecb^{n+1}$ is projected back to the global coordinate basis.
\end{enumerate}

We note again that, similarly to the numerical solution of the \hyperref[eq:GRdxdt]{GREOM} system (see previous Section \ref{sec:greom}), the half-push steps 1.\ and 3.\ could in principle be carried out in the global coordinates, without the need to project to a local inertial frame. Here, we will employ the same projection procedure in the spirit of keeping the two schemes compatible for a simultaneous use in a coupled algorithm.

\section{Coupling the GCA and the full system of equations}
\label{sec:scheme}
In this work we propose a new coupling of particle pushers that mitigates the issues related to small-scale gyration, by averaging over the gyro-motion and therefore loosening the constraints on the time time step imposed by high magnetizations. For both special- and general-relativistic simulations, the strategy consists of a dynamic switch between the \hyperref[eq:dxdt]{SREOM}/\hyperref[eq:GRdxdt]{GREOM} and the \hyperref[eq:dRdt]{SRGCA}/\hyperref[eq:GRdRdt]{GRGCA} systems of equations, based on local parameters measured at the particle position. At each time step we determine which set of equations is solved according to the following criteria:
\begin{itemize}
 \item If the particle gyro-radius is found to be smaller than a chosen fraction of the size of the computational cell $\Delta\ell$ where the particle is found, then the \hyperref[eq:dRdt]{SRGCA} or \hyperref[eq:GRdRdt]{GRGCA} system is employed for the current time-advancing step. An additional check is adopted to ensure that the value of $E$ is smaller than (a fraction of) $B$ locally, such that strong particle acceleration can be detected prior to a violation of the GCA assumptions.
 \item If the particle gyro-radius is larger than the local cell size (or if $E>B$), then the \hyperref[eq:dxdt]{SREOM} or \hyperref[eq:GRdxdt]{GREOM} system is adopted.
\end{itemize}
In numerical simulations where the fields quantities are spatially resolved on the computational grid, the minimum length over which these fields (such as $B$) can vary corresponds to the grid spacing $\sim\Delta\ell$. Therefore, by switching to the \hyperref[eq:dxdt]{SREOM}/\hyperref[eq:GRdxdt]{GREOM} system when $\rho_C\gtrsim\Delta\ell$, we also automatically filter out the case where $(B/\nabla B)<\rho_C$, i.e. the particle gyro-radius exceeds the length scale of variation of $B$, which constitutes another violation of the GCA assumptions.

The coupled strategy has the advantage that particle gyro-motion on subgrid scales is automatically averaged over in the \hyperref[eq:dRdt]{SRGCA}/\hyperref[eq:GRdRdt]{GRGCA} system, since electromagnetic fluctuations at such scales are not captured. In this way, no constraints related to accuracy loss due to fast gyration are imposed on the time step. At the same time, the motion of particles with non-negligible (larger than the cell spacing) gyro-radius is accurately captured by solving the \hyperref[eq:dxdt]{SREOM}/\hyperref[eq:GRdxdt]{GREOM} system. The switch between the two sets of equations, applied dynamically, allows for capturing particles transitioning to different physical regimes: particles with negligible gyro-motion can be accurately tracked as they gain energy, until the gyro-radius grows above the cell spacing; while energetic particles with large gyro-radii can be captured as they experience energy losses, transitioning to a regime of negligible gyro-motion.

We note that an effective cell size $\Delta\ell$ to be compared with $\rho_C$ can be reasonably defined in relation to the local cell volume, e.g.,\ $\Delta\ell=(\Delta V)^{1/n}$ for $n$-dimensional simulations. In this way, it is possible to account for the general case of non-uniform grid spacing. This criterion remains applicable in general-relativistic simulations, where the cell volume is formally defined via the proper integral
\begin{equation}
 \Delta V = \int \sqrt{\gamma} dx^1 dx^2 ... dx^n
 \label{eq:volint}
\end{equation}
evaluated over the cell's spatial extent. In the more common case of special-relativistic simulations with uniform grid spacing, one can simply take $\Delta\ell=\Delta x$.

\subsection{Special-relativistic coupled scheme}
For simulations in flat spacetimes, we apply a Boris push to solve the \hyperref[eq:dxdt]{SREOM} system, and the leapfrog pushers described in Sections \ref{sec:gcamethod1} and \ref{sec:gcamethod2} to solve the \hyperref[eq:dRdt]{SRGCA} system. The position and velocity update between consecutive time levels is therefore carried out as follows:
\begin{enumerate}
 \item Check for the validity of the GCA assumptions at the current particle location; if the \hyperref[eq:dRdt]{SRGCA} system can be employed, switch to the GCA variables $\vecR$, $\upar$, $\mu$.
 \item Update the particle position and 4-velocity with a Boris push (if solving the \hyperref[eq:dxdt]{SREOM} system) or with a GCA leapfrog step (if solving the \hyperref[eq:dRdt]{SRGCA} system) as illustrated in Sections \ref{sec:gcamethod1} or \ref{sec:gcamethod2}.
 \item After the update, store the new position and 4-velocity. If a GCA push was applied, retrieve $\vecx$ and $\vecu$ from the updated GCA variables.
\end{enumerate}

A key aspect of this coupled approach lies in making the \hyperref[eq:dRdt]{SRGCA} system and the full \hyperref[eq:dxdt]{SREOM} system compatible with each other, that is, a back-and-forth switch between the two formulations must not introduce inconsistencies. Since the GCA set of equations represents an averaging of the full equations of motion, there is an inevitable loss of information when switching to the former. Specifically, the three components of the particle velocity in three-dimensional space are reduced to the components parallel and perpendicular to the magnetic field, and information on the gyro-motion is lost. In our experiments, when switching from $\vecx$ and $\vecu$ to the GCA variables at the beginning of a time step, we first assume $\vecR\approx\vecx$ (which is a valid approximation in the limit of small gyro-radius), and then decompose the velocity as
\begin{equation}
 \vecu^{n-1/2} = \upar^{n-1/2}\vecb^{n-1/2} + \left[\vecv_E^{n-1/2} +\vecv_c(\vecR^{n-1/2},\upar^{n-1/2})\right]\Gamma(\vecR^{n-1/2},\upar^{n-1/2}) + \vecuperpg^{n-1/2},
\end{equation}
where $\upar=\vecu\cdot\vecb$, and $\vecb$, $\vecv_E$, and $\vecv_c$ are calculated with the electromagnetic fields evaluated at $\vecR^{n-1/2} = (\vecR^n+\vecR^{n-1})/2$. This directly provides $\upar$ and $\mu$ for the integration of the \hyperref[eq:dRdt]{SRGCA} system. Our assumption therefore consists of reducing the drift velocity to $\vecv_E$ and $\vecv_c$ only (which will be proven satisfactory in the following Sections). Note that, while decomposing the velocity, we also take $\Gamma(\vecR^{n-1/2},\upar^{n-1/2})$ equal to the Lorentz factor calculated from the full $\vecu^{n-1/2}$. Without this assumption, the equation above is formally implicit in $\upar^{n-1/2}$, therefore requiring an iterative procedure to obtain the GCA quantities. We verified that no differences in the results arise with the assumption above, hence we choose to avoid the iterative procedure for simplicity.

Similarly, at the end of a GCA integration step, we switch back to the full set of variables by taking the updated particle position as $\vecx\approx\vecR$. Further assumptions have to be made to reconstruct the full particle velocity from the GCA variables. The integration of the \hyperref[eq:dRdt]{SRGCA} system yields only the \textit{magnitude} of the new parallel and gyration 4-velocity, $\upar$ and $\uperpg$; the approximation introduced by the GCA becomes manifest as a lack of information on the \textit{direction} of the updated 4-velocity. Specifically, while the new direction parallel to the magnetic field is given by $\vecR$, the direction of the updated $\vecuperpg$ is unknown. We therefore choose to store $\vecb_{\perp g}^{n-1/2} = \vecuperpg^{n-1/2}/\uperpg^{n-1/2}$ at the previous time step, and reconstruct the full 4-velocity (assuming $\vecR^{n+1/2}=(\vecR^{n+1}+\vecR^n)/2$) as
\begin{equation}
 \vecu^{n+1/2} = \upar^{n+1/2}\vecb^{n+1/2} + \left[\vecv_E^{n+1/2} +\vecv_c(\vecR^{n+1/2},\upar^{n+1/2})\right]\Gamma(\vecR^{n+1/2},\upar^{n+1/2})  + \uperpg^{n+1/2}\vecb_{\perp g}^{n-1/2},
\end{equation}
hence assuming that the direction of $\vecuperpg$ (and therefore the gyration phase) is unchanged. In principle, it is possible to separately evolve the gyration phase by adding an extra equation to the \hyperref[eq:dRdt]{SRGCA} system. Here, our choice of keeping the direction of $\vecuperpg$ is justified by the fact that, if the GCA is applicable at the current integration step, the direction of $\vecuperpg$ does not play a significant role in the particle dynamics.

\subsection{General-relativistic coupled scheme}
The coupling of the \hyperref[eq:GRdxdt]{GREOM} and \hyperref[eq:GRdRdt]{GRGCA} systems (see Sections \ref{sec:greom} and \ref{sec:grgca}) can be carried out in a fashion analogous to the special-relativistic case. Notable differences lie in the fact that position and 4-velocity are now non-staggered, and that a Strang split is employed to separate the Lorentz force contribution from that of the geodesic terms. One integration step is carried out as follows:
\begin{enumerate}
 \item Check for the validity of the GCA assumptions at the current particle location; if the \hyperref[eq:GRdRdt]{GRGCA} system can be employed, switch to the GCA variables $R^i\approx x^i$, $\upar$, $\mu$.
 \item If solving the \hyperref[eq:GRdxdt]{GREOM} system, update the particle 4-velocity with a Boris push in a local inertial frame over half a time step; if solving the \hyperref[eq:GRdRdt]{GRGCA} system, update the parallel 4-velocity in a local frame over half a time step.
 \item Update the particle position and calculate the geodesic contribution to the 4-velocity (either the full $u_i$ for the \hyperref[eq:GRdxdt]{GREOM} or the parallel and drift components for the \hyperref[eq:GRdRdt]{GRGCA}) with an IMR scheme, iterating on the updated variables until convergence.
 \item Apply a second half-update to the full particle 4-velocity (if solving the \hyperref[eq:GRdxdt]{GREOM} system) or to the parallel 4-velocity (if solving the \hyperref[eq:GRdRdt]{GRGCA} system) in a local frame.
 \item After the update, store the new position and 4-velocity. If a GCA push was applied, retrieve $x^i\approx R^i$ and $u_i$ from the updated GCA variables (keeping the direction of the gyration 4-velocity fixed across a time step).
\end{enumerate}

\section{Numerical tests}
\label{sec:tests}
In this Section we test the new coupled particle pusher, assessing the accuracy and computational cost of the algorithm. We choose to run our tests in two distinct setups, with and without the effect of strong gravity. We consider static (time-invariant) electromagnetic field configurations given by PiC simulations of an isolated current sheet (in flat spacetime; see \citealt{hakobyan2019}) and of a current sheet in a BH magnetosphere (in Kerr spacetime; see \citealt{parfrey2019}). We then push test particles (which do not provide feedback to the electromagnetic fields) in these configurations, comparing the particle trajectories produced by our coupled approach with the results given by solving the \hyperref[eq:dxdt]{SREOM}/\hyperref[eq:GRdxdt]{GREOM} system with the Boris pusher. Electromagnetic fields are evaluated at the particle position $\vecx$ via interpolation from $N$ surrounding grid points, e.g.,\ $\vecB(\vecx)=\sum_N \vecB_N W(\vecx-\vecx_N)$ where $W$ is a linear shape function. Gradients of grid quantities are obtained at the particle position by differentiating $W$ with respect to $\vecx$, e.g.,\ $\nabla B(\vecx) = \sum_N B_N \nabla W(\vecx-\vecx_N)$. In order to probe the dynamics of rapidly-gyrating particles, we consider a range of values for the ratio $\rho_C/L$. This allows us to probe increasing scale separations between spatial field variations and particle gyro-motion.

\subsection{Flat spacetime: Particle motion in reconnecting current sheets}
\label{sec:testsr}
Our electromagnetic field configuration is a fully developed reconnecting current sheet. This state represents a snapshot from a PiC simulation of a magnetized pair plasma in flat spacetime (\citealt{hakobyan2019}) with \texttt{TRISTAN-MP} (\citealt{buneman1993,spitkovsky2005}). The initial configuration consists of a Harris-like profile for the magnetic field $B_y(x)=B_0\tanh(x/a)$ (\citealt{harris1962}), where $B_0$ is the asymptotic magnetic field strength and $a$ is the current sheet half-width, in a two-dimensional box of size $x\times y = 5000 d_e\times 5600 d_e$ (with grid spacing $\Delta x=0.2 d_e$), where $d_e=c/\omega_p$ is the electron skin depth and $\omega_p$ is the plasma frequency. The plasma magnetization in the ambient region, upstream of the current sheet, is adjusted via $B_0$ such that $\sigma=100$. The sheet's initial half-width is $a=4 d_e$, and no guide field is present. Along the polarity inversion region at $x=2500 d_e$, magnetic field lines break and reconnect, producing chains of plasmoids which travel in the $y$-direction. Particles travel toward the reconnection region with speed $\sim v_E$, gaining energy as they approach the X-line region. There, electric fields may grow significantly to strengths $E>B$, providing a source of strong particle acceleration.

\begin{figure}[b!]
\centering
\includegraphics[width=0.95\columnwidth, trim={9mm 0mm 7mm 0mm}, clip]{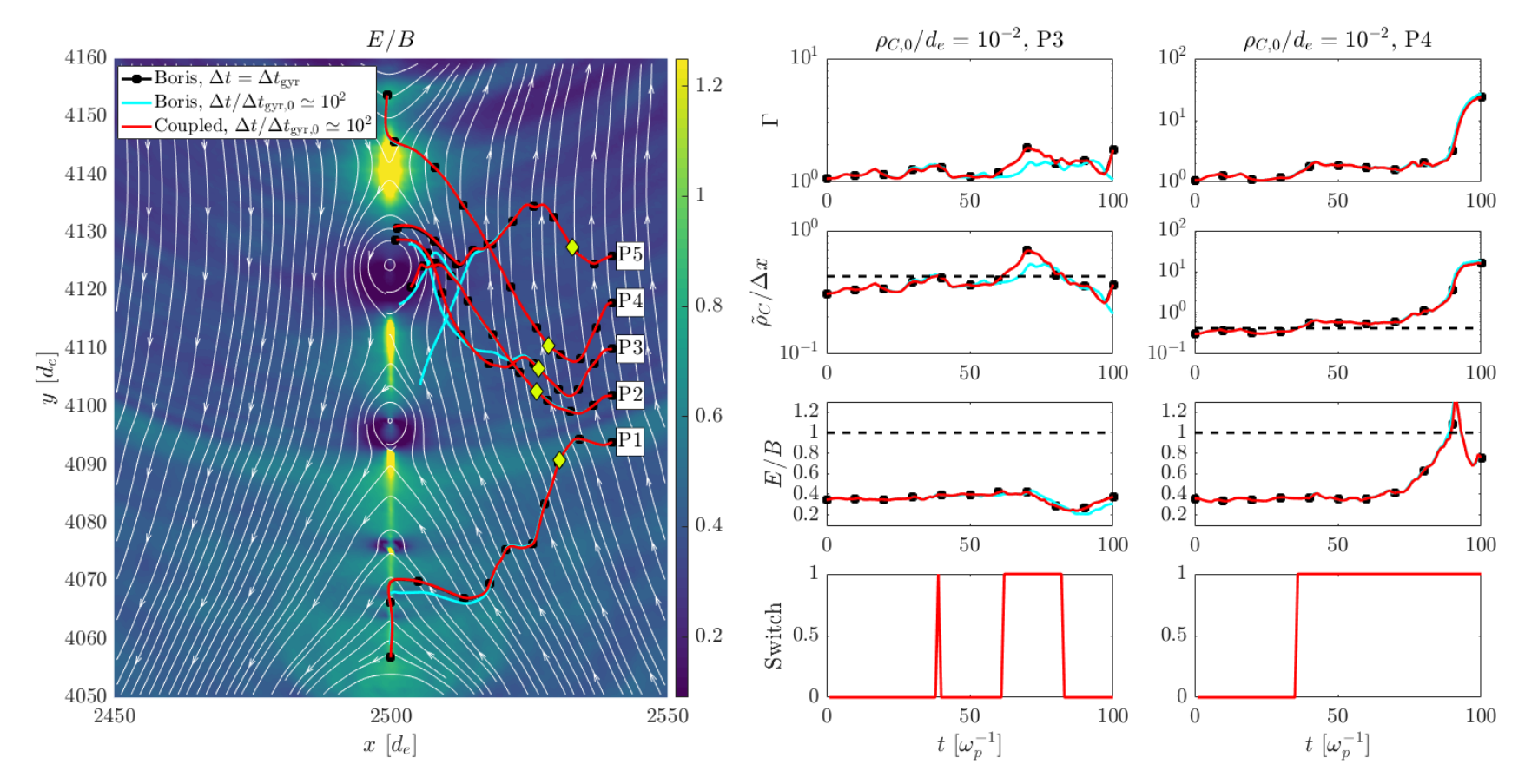}
\caption{Left: trajectories of the $\rho_{C,0}/d_e=10^{-2}$ particles produced with the Boris pusher resolving the gyration ($\Delta t=\Delta t_\mathrm{gyr}$, black lines and squares), and Boris pusher (cyan lines) and coupled pusher (red lines) with with $\Delta t/\Delta t_{\mathrm{gyr},0}\simeq10^2$. Green diamonds indicate, for the coupled pusher, the first switch between GCA and Boris schemes. Right: evolution in time of the Lorentz factor $\Gamma$ and the $\tilde{\rho}_C/\Delta x$ and $E/B$ ratios (measured at the particle position), which determine the value of the switch $S$, for particles P3 and P4. For the coupled pusher, we show the value of $S$ indicating the switch between the GCA ($S=0$) and Boris ($S=1$) algorithms. The results of the coupled scheme completely overlap with the reference Boris trajectories. The Boris scheme, instead, produces slightly diverging trajectories unless the gyration is accurately resolved at all times.}
\label{fig:CS_qm100}
\end{figure}

In this setup, we initialize 5 positrons in the ambient region at $x=2540 d_e$, $\sim 40 d_e$ away from the current sheet. The initial velocity is set to $\vecv=\vecv_E$ calculated with the value of the electromagnetic fields at the initial particle positions. The integration is carried out until $t=100\omega_p^{-1}$. The original PiC run was parametrized such that particles have an initial gyro-radius $\rho_{C,0}/d_e=10$ (and gyro-frequency $\Omega_{C,0}/\omega_p\sim 10$) away from the current sheet. In our experiments, we find that for $\rho_{C,0}/d_e\gtrsim 10^{-1}$ the Boris pusher can accurately capture the particle motion at all times, without the need for a switch to the GCA. Therefore, we explore a parameter range below this limit, $\rho_{C,0}/d_e=10^{-2},10^{-3},10^{-4}$ (corresponding to $\Omega_{C,0}/\omega_p=10^2,10^3,10^4$), in order to probe increasingly larger scale separations between spatial field variations and particle gyro-motion.

In this setup, we initialize 5 positrons in the ambient region at $x=2540 d_e$, $\sim 40 d_e$ away from the current sheet. The initial velocity is set to $\vecv=\vecv_E$ calculated with the value of the electromagnetic fields at the initial particle positions. The integration is carried out until $t=100\omega_p^{-1}$. The original PiC run was parametrized such that particles have an initial gyro-radius $\rho_{C,0}/d_e=10$ (and gyro-frequency $\Omega_{C,0}/\omega_p\sim 10$) away from the current sheet. In our experiments, we find that for $\rho_{C,0}/d_e\gtrsim 10^{-1}$ the Boris pusher can accurately capture the particle motion at all times, without the need for a switch to the GCA. Therefore, we explore a parameter range below this limit, $\rho_{C,0}/d_e=10^{-2},10^{-3},10^{-4}$ (corresponding to $\Omega_{C,0}/\omega_p=10^2,10^3,10^4$), in order to probe increasingly larger scale separations between spatial field variations and particle gyro-motion.

In order to obtain a reference solution, we run a first set of simulations with the standard Boris pusher, dynamically adapting the time step at each iteration as $\Delta t_\mathrm{gyr} = 2\pi\Omega_C^{-1}/60$, such that the particle gyration is always resolved with at least 60 steps (hence away from the current sheet, $\Delta t_\mathrm{gyr,0}\simeq 0.1\Omega_{C,0}^{-1}$). We then perform the same set of simulations with the Boris pusher, this time limiting the time step such that $\Delta t = 0.45\omega_p^{-1}$ (hence $\Delta t/\Delta t_{\mathrm{gyr},0}\simeq \Omega_{C,0}/\omega_p$). In this way, we always enforce a time step that is comparable to that employed for the PiC simulation from which the field configuration is taken (obtained from a CFL-type condition). A final set of runs with the same criterion for the time step is performed with the coupled pusher (considering both curvature and $\vecE\times\vecB$ drifts). We note that, with this choice, the ratio $\Delta t_{\mathrm{gyr},0}/\Delta t$ decreases linearly as $q/m$ increases, making the reference Boris runs progressively more expensive, at least as long as the particle is in the ambient region away from the current sheet.

To handle the switch between GCA and Boris pushers in the coupled approach, we employ the following criterion: at each time step, we measure the local value of $\tilde{\rho}_C=mc\Gamma/(|q|B)$ , i.e.\ the upper limit for a particle travelling with $\uperpg\sim\Gamma$. We then compare this value with a fraction of the grid spacing, $f_\rho$ (note that we take $\Delta\ell=\Delta x$). Additionally, we measure the local value of $E/B$ as a proxy for the magnitude of $\vecv_E$, and we compare it to a constant factor $f_E$. We therefore obtain a switch step function
\begin{equation}
 S(\tilde{\rho}_C,E/B) =
 \begin{cases}
  0 & \text{if } \tilde{\rho}_C/\Delta x < f_\rho \text{ and } E/B < f_E \\
  1 & \text{otherwise}
 \end{cases}
\end{equation}
and apply a GCA integration step if $S=0$ or a Boris push if $S=1$. In the following we find that $f_\rho=0.4$ and $f_E=1$ produce the best results for the coupled pusher.

\begin{figure}[t]
\centering
\includegraphics[width=0.95\columnwidth, trim={9mm 0mm 7mm 0mm}, clip]{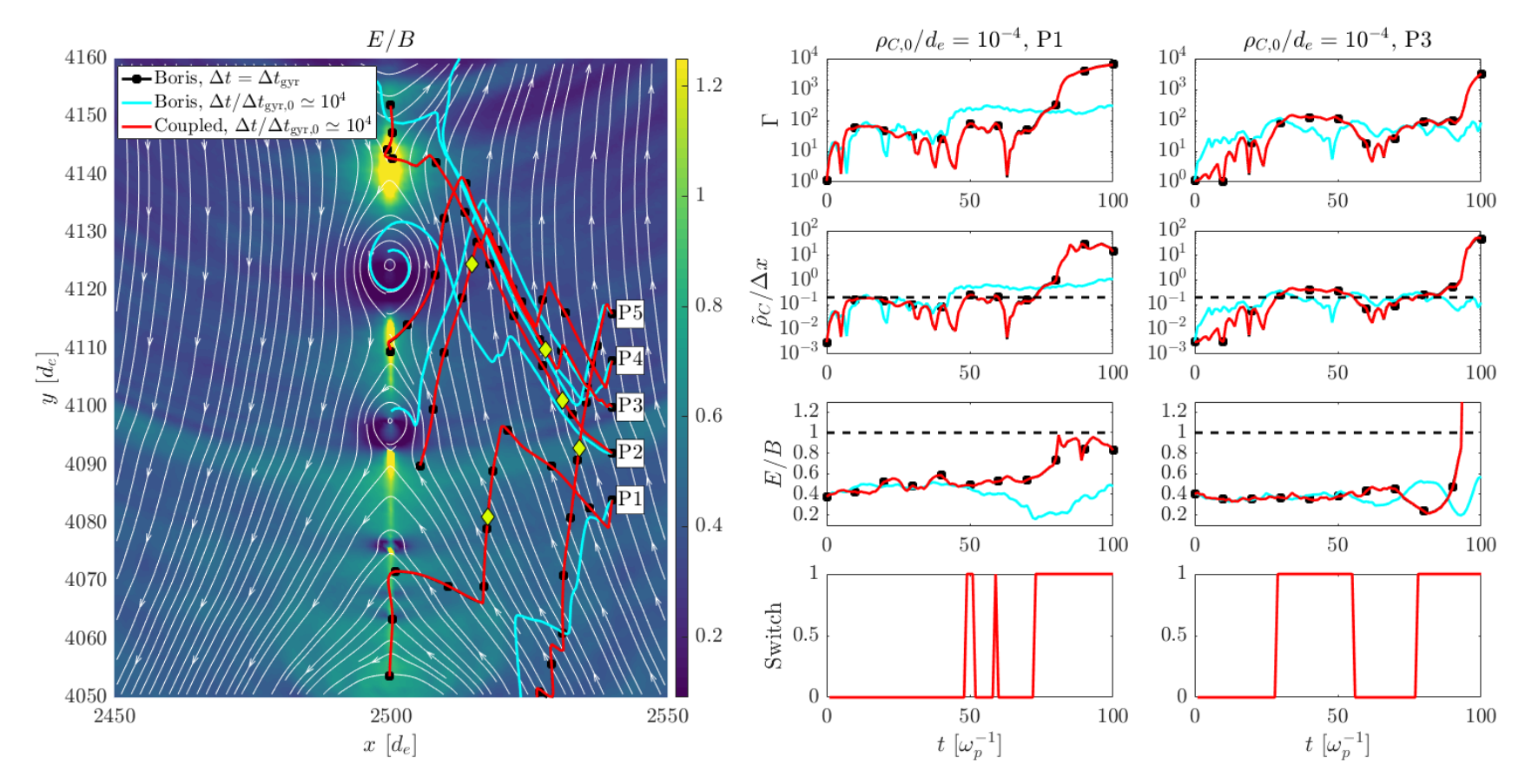}
\caption{Left: trajectories of the $\rho_{C,0}/d_e=10^{-4}$ particles produced with the Boris pusher resolving the gyration ($\Delta t=\Delta t_\mathrm{gyr}$, black lines and squares), and Boris pusher (cyan lines) and coupled pusher (red lines) with $\Delta t/\Delta t_{\mathrm{gyr},0}\simeq10^4$. Green diamonds indicate, for the coupled pusher, the first switch between GCA and Boris schemes. Right: evolution in time of the Lorentz factor $\Gamma$ and the $\tilde{\rho}_C/\Delta x$ and $E/B$ ratios (measured at the particle position), which determine the value of the switch $S$, for particles P3 and P4. For the coupled pusher, we show the value of $S$ indicating the switch between the GCA ($S=0$) and Boris ($S=1$) algorithms. The results of the coupled scheme completely overlap with the reference Boris trajectories. The Boris scheme, instead, produces substantially diverging trajectories unless the gyration is accurately resolved at all times.}
\label{fig:CS_qm10000}
\end{figure}

Our results are presented in Figures \ref{fig:CS_qm100}--\ref{fig:CS_qm10000}, where we show, for $\rho_{C,0}/d_e=10^{-2}$ and $\rho_{C,0}/d_e=10^{-4}$ and for all methods employed, the trajectory of all 5 particles (labeled at their initial positions as P1--P5) in space (left-hand panels). Green diamonds indicate, for each trajectory produced with the coupled pusher, the first switch between GCA and Boris schemes. The evolution in time of $\Gamma$, $\tilde{\rho}_C/\Delta x$, and the $E/B$ ratio for selected particles is shown in the right-hand panels. For the coupled pusher, we show the value of the switch function $S$ in time in the bottom-right panels. The switch between the GCA and Boris algorithms is indicated by the values of $\tilde{\rho}_C/\Delta x$ and $E/B$ falling above and below the $f_\rho$ and $f_E$ thresholds, respectively (dashed black lines in the $\tilde{\rho}_C/\Delta x$ and $E/B$ plots).

All particles travel from their initial position toward the current sheet, encountering spatial oscillations of the electromagnetic fields (caused by fast magnetosonic waves excited by plasmoid mergers; see \citealt{philippov2019}) along the way. As a consequence, mild oscillations in the Lorentz factor arise, before the particles reach the reconnection region. There, particles can cross regions where strong electric fields $E>B$ cause acceleration to $\Gamma\gg1$ (e.g.,\ particle P4 in Figure \ref{fig:CS_qm100} and P3 in Figure \ref{fig:CS_qm10000}. In other cases, particles can fall inside plasmoids (visible as closed, circular loops along the current sheet), where they remain trapped around the central "O-point" (a magnetic null), experiencing mild acceleration (e.g.,\ particle P3 in Figure \ref{fig:CS_qm100}).

The Boris pusher with $\Delta t/\Delta t_{\mathrm{gyr},0}\simeq \Omega_{C,0}/\omega_p$ (cyan lines), which does not accurately capture gyration at all times, shows increasingly evident discrepancies with respect to the reference Boris results with $\Delta t=\Delta t_\mathrm{gyr}$ (black lines/squares) as $\rho_{C,0}/d_e$ decreases. For $\rho_{C,0}/d_e=10^{-4}$ (i.e.\ $\Omega_{C,0}\Delta t\simeq10^4$), the inaccuracy of the Boris pusher grows so large that it produces completely diverging particle trajectories. The results of the coupled pusher with $\Delta t/\Delta t_{\mathrm{gyr},0}\simeq \Omega_{C,0}/\omega_p$ (red lines) are, instead, very consistent with the reference Boris solution in all cases. In our setup, all particles start with purely drifting motion (zero gyro-radius) and therefore their trajectories are initially captured by the GCA pusher. As they travel toward the current sheet, oscillations in the $\tilde{\rho}_C/\Delta x$ and $E/B$ ratios caused by fast magnetosonic waves can determine multiple switches between the GCA and Boris pushers. When particles reach the current sheet and cross $E>B$ regions, they can experience a significant increase in $\Gamma$ and therefore in $\tilde{\rho}_C$, which invalidates the GCA assumptions. At that point, the Boris scheme is capable of accurately resolving the particle motion, and the GCA pusher is no longer employed (e.g.,\ particle P3 in Figure \ref{fig:CS_qm10000}). Particles that approach o-points, instead, may not experience strong acceleration, such that the GCA pusher can be employed for longer times (e.g.,\ particle P3 in Figure \ref{fig:CS_qm100}).

\begin{figure}[t]
\centering
\includegraphics[width=0.95\columnwidth, trim={9mm 0mm 7mm 0mm}, clip]{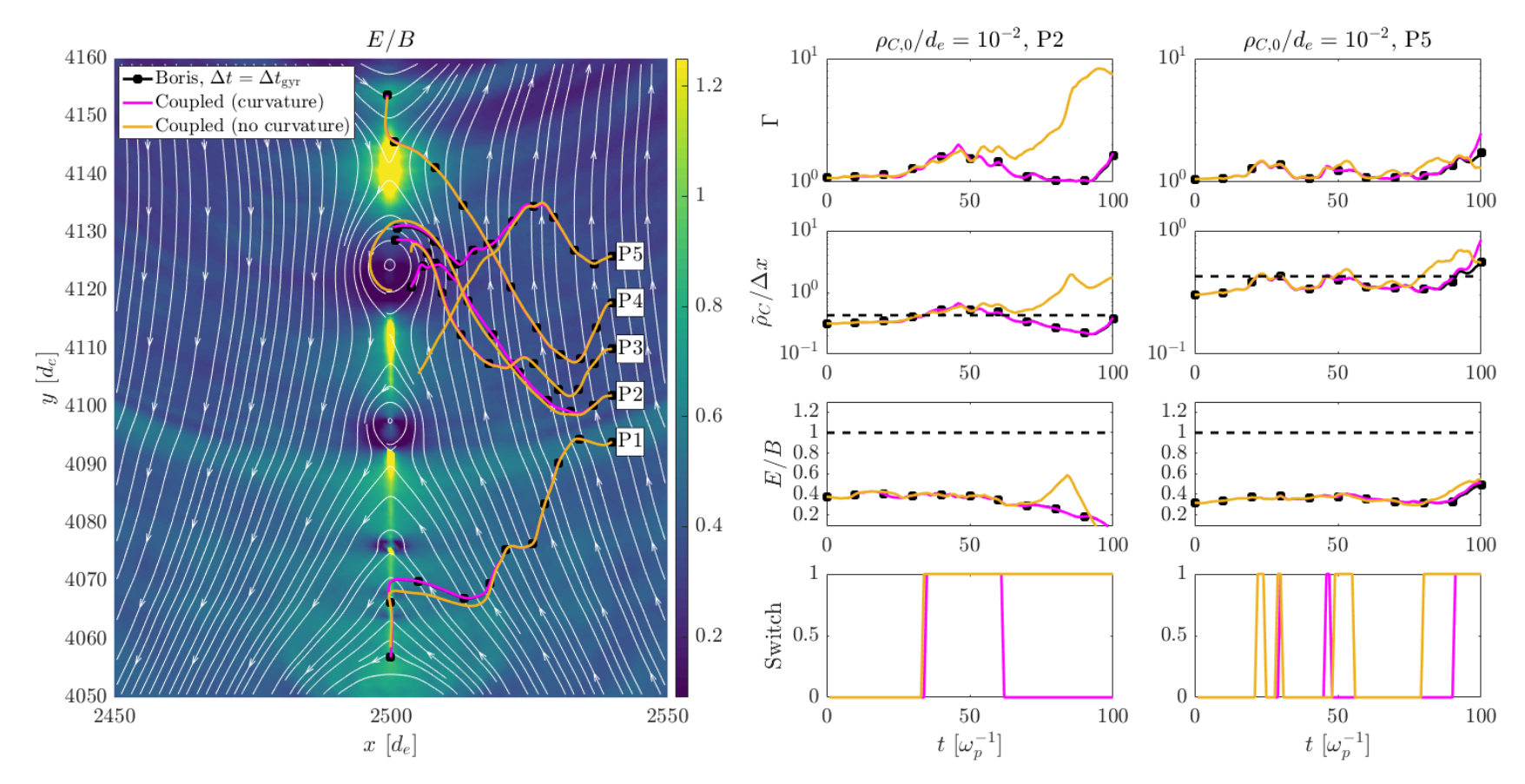}
\caption{Left: trajectories of the $\rho_{C,0}/d_e=10^{-2}$ particles produced with the Boris pusher resolving the gyration ($\Delta t=\Delta t_\mathrm{gyr}$, black lines and squares) and the coupled pushers with and without curvature drifts (magenta and orange lines respectively). Right: evolution in time of the Lorentz factor $\Gamma$ and the $\tilde{\rho}_C/\Delta x$ and $E/B$ ratios (measured at the particle position), which determine the value of the switch $S$, for particles P2 and P5. For the coupled pushers, we show the value of $S$ indicating the switch between the GCA ($S=0$) and Boris ($S=1$) algorithms. For this value of $\rho_{C,0}/d_e$, several trajectories diverge from the reference solution when neglecting curvature terms in the coupled pusher.}
\label{fig:CS_qm100_c}
\end{figure}

\begin{figure}[t]
\centering
\includegraphics[width=0.95\columnwidth, trim={9mm 0mm 7mm 0mm}, clip]{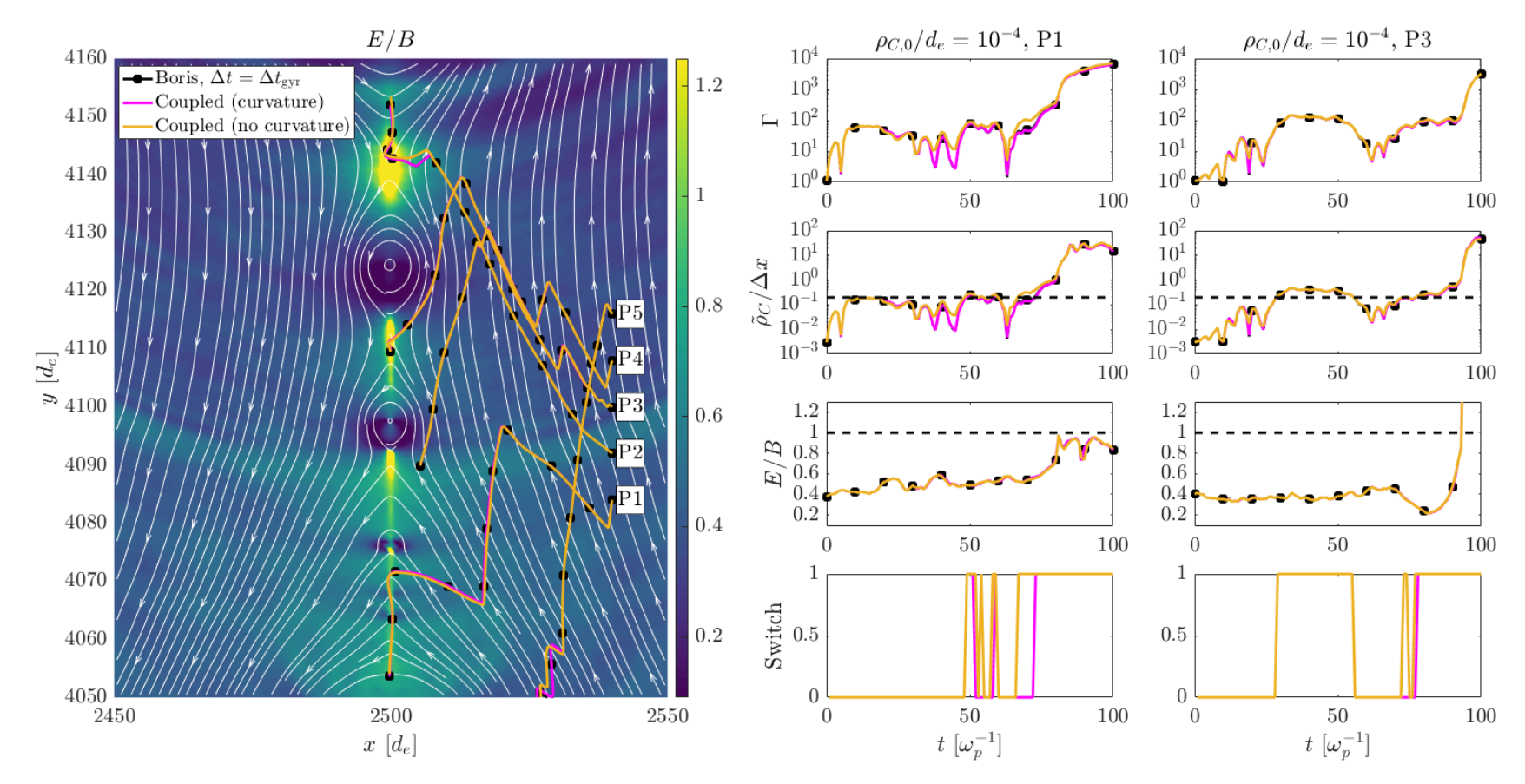}
\caption{Left: trajectories of the $\rho_{C,0}/d_e=10^{-4}$ particles produced with the Boris pusher resolving the gyration ($\Delta t=\Delta t_\mathrm{gyr}$, black lines and squares) and the coupled pushers with and without curvature drifts (magenta and orange lines respectively). Right: evolution in time of the Lorentz factor $\Gamma$ and the $\tilde{\rho}_C/\Delta x$ and $E/B$ ratios (measured at the particle position), which determine the value of the switch $S$, for particles P1 and P3. For the coupled pushers, we show the value of $S$ indicating the switch between the GCA ($S=0$) and Boris ($S=1$) algorithms. For this value of $\rho_{C,0}/d_e$, trajectories produced considering only $\vecE\times\vecB$ drifts show only minor differences with respect to those obtained when including curvature terms, indicating that $\vecE\times\vecB$ drifts are predominant.}
\label{fig:CS_qm10000_c}
\end{figure}

In a second ensemble of runs, we compare the reference Boris results with those of the coupled pusher with and without curvature terms (drift and acceleration) as illustrated in Sections \ref{sec:gcamethod1} and \ref{sec:gcamethod2}. We exemplify our results for $\rho_{C,0}/d_e=10^{-2}$ and $\rho_{C,0}/d_e=10^{-4}$ in Figures \ref{fig:CS_qm100_c}--\ref{fig:CS_qm10000_c}, where we observe that, for the lowest $\rho_{C,0}/d_e$ ratio, neglecting curvature (orange lines) has a significant effect on several particle trajectories, which can diverge from the reference solution (black lines and squares). For such trajectories, quantifying the curvature acceleration terms in equation \eqref{eq:dupardt} reveals that these are comparable in magnitude to the parallel acceleration $qE_\parallel/m$. Therefore, if curvature drifts are not taken into account, errors can arise in the particle trajectory from the solution of the GCA equations. These errors can then propagate and affect the results also when the Boris pusher is called after a switch. Instead, including curvature terms (magenta lines) reproduces the reference Boris results very accurately. As $\rho_{C,0}/d_e$ decreases, the importance of curvature effects in producing the correct trajectory decreases, and for $\rho_{C,0}/d_e=10^{-4}$ results obtained via the coupled pusher with and without curvature exhibit very small differences. This is expected, as the magnitude of the curvature terms (relative to zeroth-order terms) scales linearly with $\rho_{C}/L$, and is therefore 100 times smaller for $\rho_{C,0}/d_e=10^{-4}$. Therefore, the persisting minor differences in the trajectories where curvature drifts are not accounted for are also expected to be reduced as the magnetization is further increased. We conclude that, for strongly magnetized plasmas in the upstream of current sheets, retaining $\vecv_E$ as the only drift mechanism (as in the simple leapfrog GCA scheme of Section \ref{sec:gcamethod2}) suffices in producing accurately resolved particle trajectories with a coupled approach.

Our results show that, for large magnetizations (parametrized here via the ratio $\rho_{C,0}/d_e$) imposing small gyro-radii, the coupled pushers retains very high accuracy even with a time step that may be several orders of magnitude larger than the gyro-period. This condition occurs primarily in regions away from current sheets, where particles essentially travel in the direction of $\vecE\times\vecB$ with negligible gyro-motion. In these regions, the coupled scheme is as accurate as the Boris pusher with $\Delta t=\Delta t_\mathrm{gyr}$ (which fully resolves the gyration), but it allows for much larger (up to $\sim10^4$ times in our $\rho_{C,0}/d_e=10^{-4}$ run) time steps. Conversely, the results of the Boris pusher with such large time steps exhibit substantial inaccuracies.

\subsection{Curved spacetime: Particle motion in a black-hole magnetosphere}
\label{sec:testgr}
Our electromagnetic configuration results from an initial axisymmetric Wald solution of Maxwell's equations in Kerr spacetime (\citealt{wald1974}). In this setup, a BH of mass $M=1$ and spin $a=0.999$ (in geometrized units where $M=r_g=1$) is embedded in an external magnetic field. Magnetic field lines thread the event horizon, stretching to infinity in the direction parallel to the BH spin axis. Due to frame-dragging effects, the solution includes a nonzero electric field. The magnetic field strength at infinity is set such that away from the central object, particles have an initial gyro-radius $\rho_{C,0}/r_g\sim 10^{-3}$ (corresponding to $\Omega_{C,0}/(c/r_g)\sim 10^3$). Starting from this initial state, a general-relativistic PiC simulation (in spherical Kerr-Schild coordinates) was run, where pair plasma is continuously injected (\citealt{parfrey2019}). The simulation employed a computational grid covering the domain $0.985 r_g\leq r\leq8 r_g$, $0\leq \theta\leq\pi$ with $1280^2$ cells. The evolution of the system leads to a quasi-steady state, where the plasma is globally force-free, and a thin current sheet of length $\sim r_g$ forms in the equatorial plane. Analogously to the flat-spacetime case in the previous Section, particles entering the current sheet can be accelerated to relativistic energies by means of strong electric fields. The static (time-independent) electromagnetic fields taken from this configuration are employed to carry out our experiments.

We test our general-relativistic coupled scheme (which only includes $\vecE\times\vecB$ drifts in all cases) by initializing 5 electrons in the vicinity of the current sheet (at a distance $\sim r_g$ from the equatorial plane). The particle velocity is set equal to $\vecv_E$ at the initial particle positions. We carry out the numerical integration until $t\sim 20 r_g/c$, although most particles fall inside the event horizon well before this time.  For consistency with the parameters employed in the previous Section, we explore the range $\rho_{C,0}/r_g=10^{-2},10^{-3},10^{-4}$ ($\Omega_{C,0}/(c/r_g)= 10^2,10^3,10^4$). The choice of $\Delta t$ for each method is analogous to the flat-spacetime case, and we again obtain a reference solution by running the Boris scheme with a time step $\Delta t_\mathrm{gyr}=2\pi\Omega_C^{-1}/60$. The switch between the Strang split Boris and GCA schemes illustrated in Section \ref{sec:scheme} is handled similarly to the previous case, with a switch function $S$ assuming values of 0 or 1 corresponding to a GCA or Boris push respectively. In this case, the grid spacing $\Delta\ell$ is computed from the volume integral \eqref{eq:volint}, and we take $f_\rho=1$ and $f_E=1$.

\begin{figure}[t]
\centering
\includegraphics[width=0.95\columnwidth, trim={7mm 0mm 7mm 0mm}, clip]{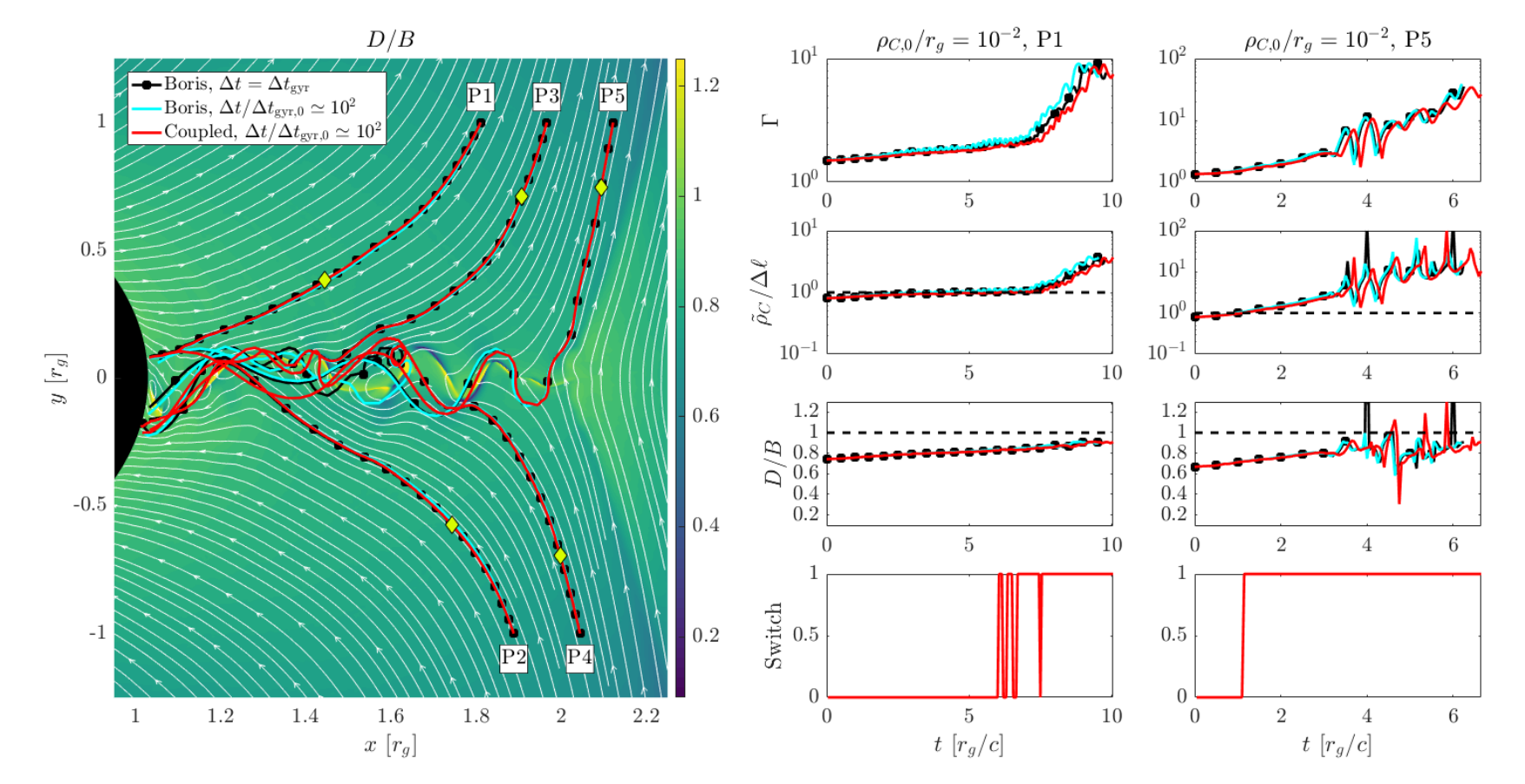}
\caption{Left: trajectories of the $\rho_{C,0}/r_g=10^{-2}$ particles produced with the Boris pusher setting $\Delta t=\Delta t_\mathrm{gyr}$ (black lines and squares), and the Boris pusher (cyan lines) and coupled pusher (red lines) with $\Delta t/\Delta t_{\mathrm{gyr},0}\simeq10^2$. Green diamonds indicate, for the coupled pusher, the first switch between the GCA and Boris schemes. Right: evolution of the Lorentz factor $\Gamma$, ratios for the switch function $\tilde{\rho}_C/\Delta\ell$ and $D/B$, and value of $S$ (for the coupled pusher; $S=0$ for GCA, $S=1$ for Boris) in time for particles P1 and P5. Consistently with the special-relativistic case, for this value of $\rho_{C,0}/r_g$ the coupled pusher introduces slight inaccuracies in the particle trajectory, due to the fact that only $\vecD\times\vecB$ drifts are included in the GCA equations.}
\label{fig:GR_qm-0.1}
\end{figure}

\begin{figure}[h]
\centering
\includegraphics[width=0.95\columnwidth, trim={7mm 0mm 7mm 0mm}, clip]{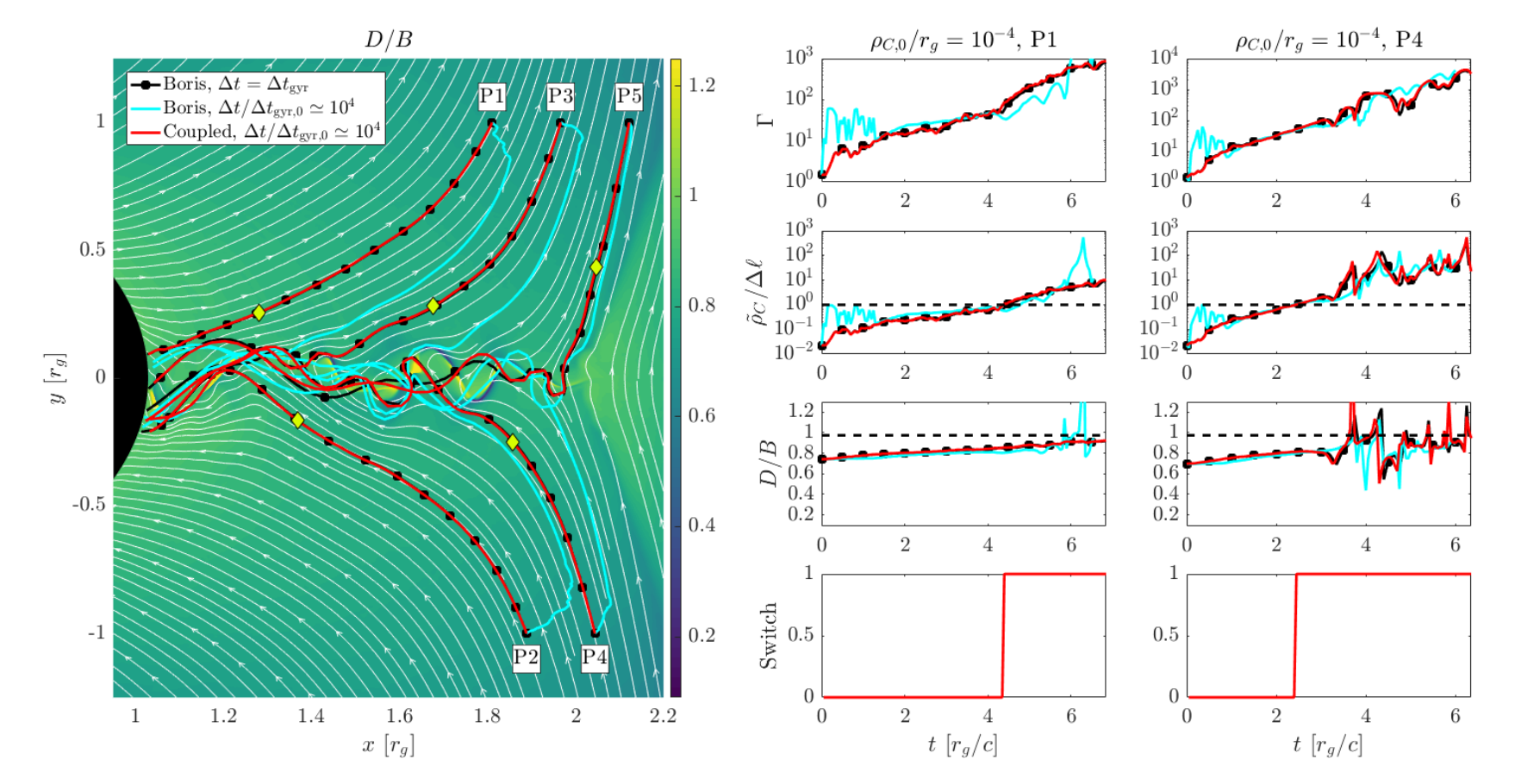}
\caption{Left: trajectories of the $\rho_{C,0}/r_g=10^{-4}$ particles produced with the Boris pusher setting $\Delta t=\Delta t_\mathrm{gyr}$ (black lines and squares), and the Boris pusher (cyan lines) and coupled pusher (red lines) with $\Delta t/\Delta t_{\mathrm{gyr},0}\simeq10^4$. Green diamonds indicate, for the coupled pusher, the first switch between the GCA and Boris schemes. Right: evolution of the Lorentz factor $\Gamma$, ratios for the switch function $\tilde{\rho}_C/\Delta\ell$ and $D/B$, and value of $S$ (for the coupled pusher; $S=0$ for GCA, $S=1$ for Boris) in time for particles P1 and P4. For this case, the coupled pusher reproduces the reference results very well, while the Boris algorithm produces large inaccuracies unless the gyro-motion is a accurately resolved with small time steps $\Delta t_\mathrm{gyr}=2\pi\Omega_C/60$ at all times.}
\label{fig:GR_qm-10}
\end{figure}

Our results are presented in Figures \ref{fig:GR_qm-0.1}--\ref{fig:GR_qm-10} for $\rho_{C,0}/r_g=10^{-2}$ and $\rho_{C,0}/r_g=10^{-4}$. We again show the trajectories for all particles with all three methods in the left-hand panels, and the time evolution of the Lorentz factor $\Gamma$, $\tilde{\rho}/\Delta\ell$, and $D/B$ ratios (with dashed black lines indicating the $f_\rho$ and $f_E$ thresholds), and value of $S$ (for the coupled pusher) for selected particles in the right-hand panels. In all cases, the particles initially drift along the $\vecD\times\vecB$ direction with zero gyro-radius. Particles that do not immediately fall into the event horizon end up in the current sheet, where they experience strong acceleration (up to $\Gamma\sim10^3$ for the smallest $\rho_{C,0}/r_g$). Differently from the flat-spacetime case, here fast magnetosonic waves from plasmoid mergers do not produce significant oscillations in $E/B$ away from the current sheet, due to a smaller scale separation in the original global GRPiC simulation. 

In agreement with the flat-spacetime test of Section \ref{sec:testsr} (see Figure \ref{fig:CS_qm100}), for $\rho_{C,0}/r_g=10^{-2}$, we observe slight discrepancies between the trajectories produced with the coupled pusher (red lines with green diamonds marking the first switch between Boris and GCA) and the reference Boris results (black lines and squares). Consistently with the flat-spacetime case, we observe a progressive improvement in the results such that, for $\rho_{C,0}/r_g = 10^{-4}$, all particle trajectories are accurately captured by the coupled pusher. This indicates a transition to a sufficiently magnetized regime where $\vecD\times\vecB$ drifts dominate the particle motion away from the current sheet. In this regime, the inclusion of only $\vecv_E$ in the GCA equations suffices in producing accurate trajectories. Conversely, as $\rho_{C,0}/r_g$ decreases the Boris scheme with $\Delta t/\Delta t_{\mathrm{gyr},0}\simeq\Omega_{C,0}/(c/r_g)$ produces progressively larger inaccuracies in the results, which are completely diverging in the $\rho_{C,0}/r_g=10^{-4}$ case. We note that, for the coupled pusher, the thresholds $f_\rho=f_E=1$ work well to activate the switch and to obtain accurate trajectories. For large charge-to-mass ratios, these thresholds act such that the GCA equations are employed until shortly before the particles reach the current sheet, where strong acceleration sharply raises the $\tilde{\rho}_C/\Delta\ell$ ratio. At that point, the Boris pusher is capable of fully resolving the particle motion, and numerical errors caused by overstepping the gyro-period are avoided.

In a final set of runs, we consider particle motion in the ambient region (outside of the region where the BH rotation twists magnetic field lines, $r\sin\theta > 2r_g$), where highly magnetized particles travel along (essentially) straight field lines with almost constant velocity and negligible gyro-motion. For this test, we initialize 4 electrons such that $\rho_{C,0}/r_g=10^{-1},10^{-2},10^{-3},10^{-4}$ (i.e.\ $\Omega_{C,0}/(c/r_g)=10^{1},10^{2},10^{3},10^{4}$). All particles are initially placed at $r=5.5r_g, \theta=\pi/2$ with zero velocity, and we carry out the integration until $t=20 r_g/c$. The thresholds $f_\rho=f_E=1$ are employed for the coupled pusher.

\begin{figure}[h]
\centering
\includegraphics[width=1\columnwidth, trim={35mm 0mm 17mm 0mm}, clip]{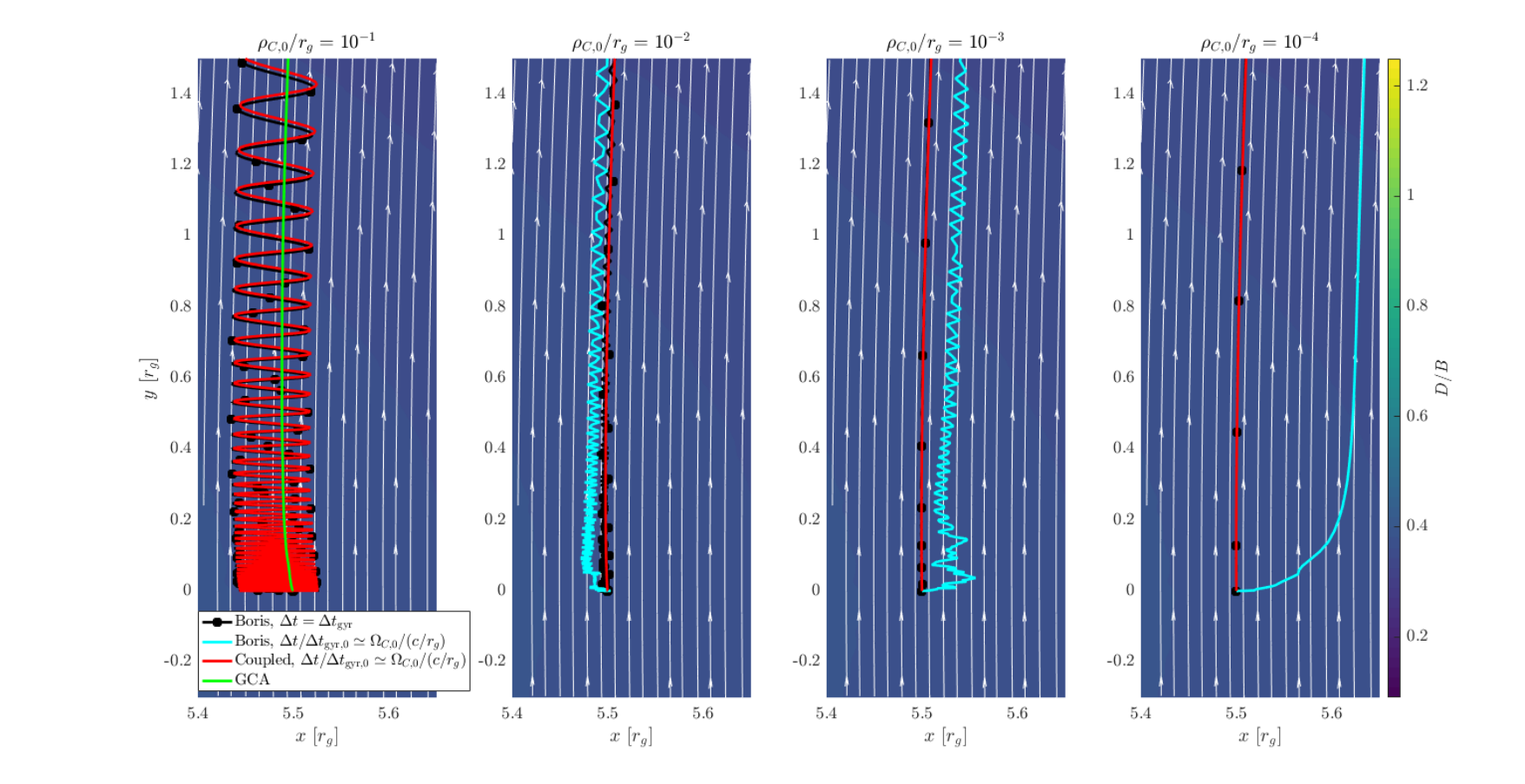}
\caption{Trajectories of 4 particles ($\rho_{C,0}/r_g=10^{-1}$, $10^{-2}$, $10^{-3}$, $10^{-4}$) in the ambient region away from the central BH. The Boris pusher with $\Delta t=\Delta t_\mathrm{gyr}$ (black lines and squares) provides reference results where, for large magnetizations (i.e.\ $\rho_{C,0}/r_g<10^{-1}$), particles travel along straight magnetic field lines with negligible gyro-motion. When $\Delta t/\Delta t_{\mathrm{gyr},0}\simeq\Omega_{C,0}/(c/r_g)$, the Boris pusher (cyan lines) remains accurate only as long as $\rho_{C,0}/r_g\ge10^{-1}$; for larger magnetizations, the small-scale gyration is not accurately resolved at all times, and the scheme introduces large inaccuracies in the trajectories, which substantially diverge from the reference solutions. In the coupled pusher (red lines), instead, small-scale gyration always triggers a call to the GCA solver only, and the trajectories perfectly overlap with the reference results. For the case $\rho_{C,0}/r_g=10^{-1}$, where the gyro-radius remains well above the grid spacing, we overplot the guiding center trajectory (green line) for reference.}
\label{fig:GR_ambient}
\end{figure}

The results are presented in Figure \ref{fig:GR_ambient}, where we show part of the reference Boris trajectories (black lines and squares) obtained by accurately resolving the gyro-motion, and the results of the Boris and coupled schemes with $\Delta t/\Delta t_{\mathrm{gyr},0}\simeq\Omega_{C,0}/(c/r_g)$ (cyan and red lines respectively). For $\rho_{C,0}/r_g=10^{-1}$, the gyro-radius is well above the grid spacing, such that the gyro-motion is clearly visible. In this case, the Boris scheme with $\Delta t/\Delta t_{\mathrm{gyr},0}\simeq\Omega_{C,0}/(c/r_g)$ can accurately resolve the gyration, hence reproducing the reference solution very well. The coupled scheme performs equally well, never switching to the GCA equations, and overlaps the Boris solution perfectly. As a reference, for this case we overplot the guiding center trajectory produced by integrating the \hyperref[eq:GRdRdt]{GRGCA} equations (green line). As the magnetization increases such that $\rho_{C,0}/r_g\le10^{-2}$, the gyro-radius progressively shrinks below the grid spacing, therefore requiring extremely small time steps to accurately resolve the gyration. In the Boris scheme with $\Delta t/\Delta t_{\mathrm{gyr},0}\simeq\Omega_{C,0}/(c/r_g)$ the gyro-period becomes significantly overstepped, introducing progressively larger inaccuracies which result in completely diverging trajectories. In the coupled scheme, instead, the subgrid gyro-motion is detected via the $f_\rho$ threshold, such that the Boris pusher is never employed, and the trajectory is calculated via the GCA equations at all times. As a consequence, the results of the coupled pusher reproduce the reference trajectories extremely well also for large magnetizations, with no signs of the inaccuracies that characterize the under-resolved Boris results. The case $\rho_{C,0}/r_g=10^{-4}$ does not represent a limiting value for the applicability of the coupled scheme: in our experiments, we further pushed the scale separation to values $\rho_{C,0}/r_g<10^{-7}$, with completely analogous results, indicating that the coupled scheme is capable of retaining high accuracy even for gyro-motion on extremely small scales.

Our tests confirm that, for strongly magnetized plasmas and even for scenarios where strong gravity plays an important role, the coupled pusher can produce reliable physical results, capturing the motion of particles with gyro-radius much smaller the grid spacing. Conversely, the standard Boris pusher produces large inaccuracies that can only be mitigated by taking much smaller time steps (we observe large inaccuracies for $\Omega_{C,0}\Delta t\gtrsim10$), increasing the computational requirements dramatically (up to a factor $\sim r_g/\rho_{C,0}$ larger than the cost of the coupled pusher).

\section{Conclusions}
\label{sec:conclusions}

Plasma in collisionless magnetospheres of NSs and BHs is usually highly magnetized, such that plasma particles gyrate with gyro-frequencies that are orders of magnitude larger than the frequency of rotation of the compact object. In this regime, particle gyration becomes negligible, and the particle motion practically reduces to that of the guiding center. However, in dissipation regions, e.g., magnetic reconnection layers where the local magnetic field strength goes to zero, particles can accelerate and gain significant transverse momentum resulting in non-negligible gyro-radii. These conditions put stringent constraints on the time step of numerical algorithms for plasma simulations of compact-object magnetospheres, and the system's physical parameters are in practice typically rescaled to artificially decrease the scale separation.

To mitigate this problem we designed a novel coupled particle pusher that can accurately resolve particle dynamics both in highly magnetized plasmas and in acceleration regions where the local magnetic field strength decreases. The algorithm dynamically switches between solving the full equations of motion with a Boris scheme, and the guiding center equations of motion with a newly developed semi-implicit leap-frog scheme, depending on the magnetization of the plasma particles. In magnetized regions we solve only for the guiding center motion of the orbit, averaging over the particle's gyration. In this way the restriction on the time step due to very small gyro-frequency is alleviated, while the particle dynamics is accurately resolved both for large and small gyro-radii. This is appropriate for example for typical field strengths near pulsars and magnetars where synchrotron losses very quickly push particles to the zeroth Landau level.

We tested the new algorithm for typical plasma parameters in compact-object magnetospheres. We considered two cases, namely i) an isolated reconnecting current sheet in magnetized plasma, and ii) a reconnection layer that formed in the magnetized magnetosphere of a BH. We compared the motion of charged particles with our new coupled pusher to a reference solution of the full equations of motion where the gyro-frequency is fully resolved, for a range of particle magnetizations. In this way, we probed increasingly larger scale separations between the particle gyro-motion and the length scales of variation of the electromagnetic fields (which are resolved on the computational grid). We find that the new coupled scheme can reproduce the reference results very well for the large scale separations we considered here, as applicable in magnetospheres of compact objects. We also show that for particles moving in the magnetized upstream (away from current sheets) all drifts associated with field curvature can be safely neglected for the strong magnetic fields near compact objects. Our particle pusher is presented such that it can be readily implemented in existing PiC algorithms, both in flat spacetime and in a dedicated covariant formulation for curved spacetime. The scheme can also be used to trace test particles in magnetohydrodynamic simulations, for example to study particle acceleration due to resistive relativistic reconnection in compact-object magnetospheres, accretion disks, coronae, and jets (\citealt{ripperda2018b,ripperda2019a,ripperda2019b,ripperda2020}). In future applications we will focus on global simulations of particle acceleration in current sheets that form in highly magnetized compact-object magnetospheres. In these environments, when particles accelerate to high energies, they are typically subject to radiative (e.g.\ synchrotron) losses. In order to probe the fast-cooling regime of energetic particles in the current sheet (\citealt{cerutti2016,philippovspitkovsky2018}), previous PiC simulations of pulsar magnetospheres required to take very small time steps such that the Larmor frequency is resolved everywhere, including regions near the stellar surface where gyration can become extremely fast\footnote{This is necessary because accurate accounting of the synchrotron losses requires resolving the Larmor frequency in time. Under-resolving Larmor gyrations in time, while keeping the same synchrotron cooling strength, results in numerical instabilities.}. Our new coupled pusher removes this limitation, as only high-energy particles, pushed with a Boris algorithm, are subject to synchrotron cooling. Note that our formulation of the GCA equations implicitly assumes that all GCA particles are indeed in a state of near-zero magnetic moment  (which in realistic situations can occur e.g., due to radiative cooling). In the unlikely case where the magnetic moment remains non-negligible, but at the same time the gyro-radius is microscopic, the scheme could suffer from significant inaccuracies. Therefore, care should be taken in applying the algorithm to these specific situations, which however we do not foresee in the applications we target.

As a way forward to truly multi-scale simulations of relativistic magnetospheres of compact objects, a particularly attractive approach for these magnetized plasmas is to couple a PiC method with a force-free description, where magnetic fields are considered so strong that the inertia of the plasma can be ignored; highly magnetized ideal (i.e., where $\vecE\cdot\vecB=0$) parts of the computational domain can be treated with a simple force-free algorithm, while particle acceleration (and the associated feedback on electromagnetic fields) in dissipation regions, like current sheets, is accurately captured by a PiC method. While our coupled GCA-Boris approach alleviates constraints on the time step based on the gyro-radius and gyro-frequency, the skin depth and associated plasma frequency still require to be resolved in explicit PiC simulations. Ultimately, an implicit PiC approach can alleviate constraints to resolve the plasma frequency when dynamics take place on scales much larger than the skin depth (\citealt{bacchini2019b}). Such an approach can be combined with an adaptive or stretched mesh, that refines particularly in dissipation regions, while remaining coarse in the magnetized upstream. We will work on developing these new numerical methods in the future.

\section*{Acknowledgements}
The authors wish to thank Hayk Hakobyan for providing the PiC data used for part of the tests presented in this work. The computational resources and services used in this work were provided by the VSC (Flemish Supercomputer Center), funded by the Research Foundation -- Flanders (FWO) and the Flemish Government -- department EWI. FB is supported by a Junior PostDoctoral Fellowship (grant number 12ZW220N) from Research Foundation -- Flanders (FWO). BR is supported by a Joint Princeton/Flatiron Postdoctoral Fellowship. Research at the Flatiron Institute is supported by the Simons Foundation, which also supported KP.

\bibliographystyle{apalike}

\begin{thebibliography}{}
\expandafter\ifx\csname natexlab\endcsname\relax\def\natexlab#1{#1}\fi

\bibitem[{Arnowitt {et~al.}(1959)Arnowitt, Deser, \& Misner}]{arnowitt1959}
Arnowitt, R., Deser, S., \& Misner, C. 1959, Phys. Rev., 116, 5

\bibitem[{{Bacchini} {et~al.}(2019{\natexlab{a}}){Bacchini}, Amaya, \&
  Lapenta}]{bacchini2019b}
{Bacchini}, F., Amaya, J., \& Lapenta, G. 2019{\natexlab{a}}, J. Phys: Conf.
  Ser., 1225, 012011

\bibitem[{Bacchini {et~al.}(2018)Bacchini, Ripperda, Chen, \&
  Sironi}]{bacchini2018a}
Bacchini, F., Ripperda, B., Chen, A., \& Sironi, L. 2018, ApJS, 237, 6

\bibitem[{{Bacchini} {et~al.}(2019{\natexlab{b}}){Bacchini}, {Ripperda},
  {Porth}, \& {Sironi}}]{bacchini2019a}
{Bacchini}, F., {Ripperda}, B., {Porth}, O., \& {Sironi}, L.
  2019{\natexlab{b}}, ApJS, 240, 40

\bibitem[{Beklemishev \& Tessarotto(1999)}]{beklemishevtessarotto1999}
Beklemishev, A., \& Tessarotto, M. 1999, PoP, 6, 4487

\bibitem[{Beklemishev \& Tessarotto(2004)}]{beklemishevtessarotto2004}
---. 2004, A\&A, 428, 1

\bibitem[{Birn {et~al.}(2004)Birn, Thomsen, \& Hesse}]{birn2004}
Birn, J., Thomsen, M., \& Hesse, M. 2004, PoP, 11, 1825

\bibitem[{Boris(1970)}]{boris1970}
Boris, J. 1970, Proceedings of the Fourth Conference on Numerical Simulations
  of Plasmas (Naval Research Laboratory, Washington DC, 1970), p. 3.

\bibitem[{Buneman(1993)}]{buneman1993}
Buneman, O. 1993, Computer Space Plasma Physics, ed. H. Matsumoto and Y. Omura
  (Tokyo: Terra Scientific), 67

\bibitem[{{Cerutti} {et~al.}(2016){Cerutti}, {Philippov}, \&
  {Spitkovsky}}]{cerutti2016}
{Cerutti}, B., {Philippov}, A., \& {Spitkovsky}, A. 2016, MNRAS, 457, 3

\bibitem[{Chen \& Beloborodov(2017)}]{chenbeloborodov2014}
Chen, A., \& Beloborodov, A. 2017, ApJ, 844, 133

\bibitem[{Chen \& Yuan(2019)}]{chenyuan2019}
Chen, A., \& Yuan, Y. 2019, arXiv:1908.06919

\bibitem[{Comisso \& Sironi(2018)}]{comissosironi2018}
Comisso, L., \& Sironi, L. 2018, PRL, 121, 25

\bibitem[{Comisso \& Sironi(2019)}]{comissosironi2019}
---. 2019, ApJ, 886, 122

\bibitem[{Cremaschini {et~al.}(2008)Cremaschini, Tessarotto, Nicolini, \&
  Beklemishev}]{cremaschini2008}
Cremaschini, C., Tessarotto, M., Nicolini, P., \& Beklemishev, A. 2008, AIP
  Conference Proceedings, 1084, 1

\bibitem[{Crinquand {et~al.}(2020)Crinquand, Cerutti, Philippov, Parfrey, \&
  Dubus}]{crinquand2020}
Crinquand, B., Cerutti, B., Philippov, A., Parfrey, K., \& Dubus, G. 2020, PRL,
  124, 145101

\bibitem[{Gordovskyy \& Browning(2011{\natexlab{a}})}]{gordovskyybrowning2011a}
Gordovskyy, M., \& Browning, P. 2011{\natexlab{a}}, Sol. Phys, 277, 2

\bibitem[{Gordovskyy \& Browning(2011{\natexlab{b}})}]{gordovskyybrowning2011b}
---. 2011{\natexlab{b}}, ApJ, 729, 2

\bibitem[{Gordovskyy {et~al.}(2014)Gordovskyy, Browning, Kontar, \&
  Bian}]{gordovskyy2014}
Gordovskyy, M., Browning, P., Kontar, E., \& Bian, N. 2014, A\&A, 561, 72

\bibitem[{Gordovskyy {et~al.}(2010)Gordovskyy, Browning, \&
  Vekstein}]{gordovskyy2010}
Gordovskyy, M., Browning, P., \& Vekstein, G. 2010, ApJ, 720, 2

\bibitem[{Guo {et~al.}(2014)Guo, Li, Daughton, \& Liu}]{guo2014}
Guo, F., Li, H., Daughton, W., \& Liu, Y. 2014, PRL, 113, 155005

\bibitem[{Hairer {et~al.}(2006)Hairer, Lubich, \& Wanner}]{hairer}
Hairer, E., Lubich, C., \& Wanner, G. 2006, Geometric Numerical Integration
  (Zhejiang Publishing United Group, Zhejiang Science and Technology Publishing
  House, Hangzhou)

\bibitem[{Hakobyan {et~al.}(2019)Hakobyan, Philippov, \&
  Spitkovsky}]{hakobyan2019}
Hakobyan, H., Philippov, A., \& Spitkovsky, A. 2019, ApJ, 877, 53

\bibitem[{Harris(1962)}]{harris1962}
Harris, E. 1962, Il Nuovo Cimento, 23,115

\bibitem[{{Kalapotharakos} {et~al.}(2018){Kalapotharakos}, {Brambilla},
  {Timokhin}, {Harding}, \& {Kazanas}}]{kalapotharakos2018}
{Kalapotharakos}, C., {Brambilla}, G., {Timokhin}, A., {Harding}, A., \&
  {Kazanas}, D. 2018, ApJ, 857, 1

\bibitem[{Leroy {et~al.}(2019)Leroy, Ripperda, \& Keppens}]{leroy2019}
Leroy, M. H.~J., Ripperda, B., \& Keppens, R. 2019, JGR:SP, 124, 8

\bibitem[{Levinson \& Cerutti(2018)}]{levinsoncerutti2018}
Levinson, A., \& Cerutti, B. 2018, A\&A, 616, A184

\bibitem[{Northrop(1961)}]{northrop1961}
Northrop, T. 1961, Annals of Physics, 15, 79

\bibitem[{Northrop(1963)}]{northrop1963}
---. 1963, Reviews of Geophysics and Space Physics, 1, 283

\bibitem[{{Parfrey} {et~al.}(2019){Parfrey}, {Philippov}, \&
  {Cerutti}}]{parfrey2019}
{Parfrey}, K., {Philippov}, A., \& {Cerutti}, B. 2019, PRL, 122, 035101

\bibitem[{Philippov {et~al.}(2020)Philippov, Timokhin, \&
  Spitkovsky}]{philippov2020}
Philippov, A., Timokhin, A., \& Spitkovsky, A. 2020, arXiv:2001.02236

\bibitem[{Philippov {et~al.}(2019)Philippov, Uzdensky, Spitkovsky, \&
  Cerutti}]{philippov2019}
Philippov, A., Uzdensky, D., Spitkovsky, A., \& Cerutti, B. 2019, ApJL, 876, L6

\bibitem[{Philippov {et~al.}(2015)Philippov, Cerutti, Tchekhovskoy, \&
  Spitkovsky}]{philippov2015b}
Philippov, A.~A., Cerutti, B., Tchekhovskoy, A., \& Spitkovsky, A. 2015, ApJL,
  815, L19

\bibitem[{Philippov \& Spitkovsky(2018)}]{philippovspitkovsky2018}
Philippov, A.~A., \& Spitkovsky, A. 2018, ApJ, 855, 94

\bibitem[{Pinto {et~al.}(2016)Pinto, Gordovskyy, Browning, \&
  Vilmer}]{pinto2016}
Pinto, R., Gordovskyy, M., Browning, P., \& Vilmer, M. 2016, A\&A, 585, 159

\bibitem[{Ripperda {et~al.}(2020)Ripperda, Bacchini, \&
  Philippov}]{ripperda2020}
Ripperda, B., Bacchini, F., \& Philippov, A. 2020, arXiv:2003.04330

\bibitem[{Ripperda {et~al.}(2018)Ripperda, Bacchini, Teunissen, Xia, Porth,
  Sironi, Lapenta, \& Keppens}]{ripperda2018a}
Ripperda, B., Bacchini, F., Teunissen, J., {et~al.} 2018, ApJS, 235, 21

\bibitem[{Ripperda {et~al.}(2019{\natexlab{a}})Ripperda, Porth, \&
  Keppens}]{ripperda2019a}
Ripperda, B., Porth, O., \& Keppens, R. 2019{\natexlab{a}}, Journal of Physics:
  Conference Series, 1225, 01201

\bibitem[{{Ripperda} {et~al.}(2018){Ripperda}, {Porth}, {Sironi}, \&
  {Keppens}}]{ripperda2018b}
{Ripperda}, B., {Porth}, O., {Sironi}, L., \& {Keppens}, R. 2018, MNRAS, 485, 1

\bibitem[{Ripperda {et~al.}(2017{\natexlab{a}})Ripperda, Porth, Xia, \&
  Keppens}]{ripperda2017a}
Ripperda, B., Porth, O., Xia, C., \& Keppens, R. 2017{\natexlab{a}}, MNRAS,
  467, 3

\bibitem[{Ripperda {et~al.}(2017{\natexlab{b}})Ripperda, Porth, Xia, \&
  Keppens}]{ripperda2017b}
---. 2017{\natexlab{b}}, MNRAS, 471, 3

\bibitem[{Ripperda {et~al.}(2019{\natexlab{b}})Ripperda, Bacchini, Porth, Most,
  Olivares, Nathanail, Rezzolla, Teunissen, \& Keppens}]{ripperda2019b}
Ripperda, B., Bacchini, F., Porth, O., {et~al.} 2019{\natexlab{b}}, ApJS, 244,
  10

\bibitem[{Rosdahl \& Galsgaard(2009)}]{rosdhalgalsgaard2009}
Rosdahl, K., \& Galsgaard, K. 2009, A\&A, 511, 73

\bibitem[{Sironi \& Spitkovsky(2009{\natexlab{a}})}]{sironispitkovsky2009a}
Sironi, L., \& Spitkovsky, A. 2009{\natexlab{a}}, ApJ, 698, 2

\bibitem[{Sironi \& Spitkovsky(2009{\natexlab{b}})}]{sironispitkovsky2009b}
---. 2009{\natexlab{b}}, ApJL, 707, 1

\bibitem[{Sironi \& Spitkovsky(2014)}]{sironispitkovsky2014}
---. 2014, ApJL, 783, L21

\bibitem[{Spitkovsky(2005)}]{spitkovsky2005}
Spitkovsky, A. 2005, AIP Conf. Proc. 801, Astrophysical Sources of High Energy
  Particles and Radiation, ed. T. Bulik, B. Rudak, and G. Madejski (Melville,
  NY: AIP), 345

\bibitem[{Spitkovsky(2008)}]{spitkovsky2008}
---. 2008, ApJL, 682, 1

\bibitem[{Threlfall {et~al.}(2017)Threlfall, Neukirch, \&
  Parnell}]{threlfall2017}
Threlfall, J., Neukirch, T., \& Parnell, C. 2017, Sol. Phys., 292, 45

\bibitem[{Vandervoort(1960)}]{vandervoort1960}
Vandervoort, P. 1960, Ann. Phys., 10, 401

\bibitem[{Wald(1974)}]{wald1974}
Wald, R.~M. 1974, Phys. Rev. D, 10, 1680

\bibitem[{Werner \& Uzdensky(2017)}]{werneruzdensky2017}
Werner, G., \& Uzdensky, D. 2017, ApJL, 843, 2

\bibitem[{Werner {et~al.}(2018)Werner, Uzdensky, Begelman, Cerutti, \&
  Nalewajko}]{werner2018}
Werner, G., Uzdensky, D., Begelman, M., Cerutti, B., \& Nalewajko, K. 2018,
  MNRAS, 473, 4

\bibitem[{Zhdankin {et~al.}(2019)Zhdankin, Uzdensky, Werner, \&
  Begelman}]{zhdankin2019}
Zhdankin, V., Uzdensky, D., Werner, G., \& Begelman, M. 2019, PRL, 122, 5

\bibitem[{Zhdankin {et~al.}(2018)Zhdankin, Uzdensky, Werner, \&
  Begelman}]{zhdankin2018}
Zhdankin, V., Uzdensky, D.~A., Werner, G.~R., \& Begelman, M.~C. 2018, MNRAS,
  474, 2

\bibitem[{Zhou {et~al.}(2015)Zhou, Buechner, Barta, Gan, \& Liu}]{zhou2015}
Zhou, X., Buechner, J., Barta, M., Gan, W., \& Liu, S. 2015, ApJ, 815, 6

\bibitem[{Zhou {et~al.}(2016)Zhou, Buechner, Barta, Gan, \& Liu}]{zhou2016}
---. 2016, ApJ, 827, 94

\end{thebibliography}

\label{lastpage}
\end{document}